\documentclass[12pt]{article}

\usepackage{amsmath, amsthm, amssymb}
\usepackage{setspace}
\usepackage{multirow}
\usepackage{graphicx}
\usepackage{apacite}
\usepackage{natbib}
\usepackage{enumitem}
\usepackage{booktabs}

\newcommand{\ua}       {\mbox{\boldmath$a$}}

\newcommand{\uC}       {\mbox{\boldmath$C$}}

\newcommand{\uD}       {\mbox{\boldmath$D$}}
\newcommand{\ud}       {\mbox{\boldmath$d$}}
\newcommand{\uE}       {\mbox{\boldmath$E$}}

\newcommand{\uH}       {\mbox{\boldmath$H$}}

\newcommand{\uI}       {\mbox{\boldmath$I$}}

\newcommand{\uS}       {\mbox{\boldmath$S$}}

\newcommand{\uW}       {\mbox{\boldmath$W$}}

\newcommand{\uX}       {\mbox{\boldmath$X$}}
\newcommand{\ux}       {\mbox{\boldmath$x$}}

\newcommand{\uy}       {\mbox{\boldmath$y$}}
\newcommand{\uZ}       {\mbox{\boldmath$Z$}}
\newcommand{\uz}       {\mbox{\boldmath$z$}}

\newcommand{\udelta}       {\mbox{\boldmath$\delta$}}

\newcommand{\uEta}       {\mbox{\boldmath$H$}}

\newcommand{\utheta}       {\mbox{\boldmath$\theta$}}
\newcommand{\uTheta}       {\mbox{\boldmath$\Theta$}}

\newcommand{\ulambda}       {\mbox{\boldmath$\lambda$}}
\newcommand{\umu}       {\mbox{\boldmath$\mu$}}
\newcommand{\unu}       {\mbox{\boldmath$\nu$}}
\newcommand{\uxi}       {\mbox{\boldmath$\xi$}}

\newcommand{\uSigma}       {\mbox{\boldmath$\Sigma$}}

\newcommand{\uzero}       {\mbox{\boldmath$0$}}
\newcommand{\uone}       {\mbox{\boldmath$1$}}

\DeclareMathOperator{\tr}{tr}

\title{Bayesian density regression for discrete outcomes}

\author{Georgios Papageorgiou\\
	Department of Economics, Mathematics and Statistics\\
	Birkbeck, University of London, UK\\
	g.papageorgiou@bbk.ac.uk}

\begin{document}
\maketitle
	
\begin{center}
\emph{Abstract}
\end{center}
We develop Bayesian models for density regression with emphasis on discrete outcomes. 
The problem of density regression is approached by considering methods for multivariate 
density estimation of mixed scale variables, and obtaining conditional densities from the 
multivariate ones. The approach to multivariate mixed scale outcome density estimation that we describe 
represents discrete variables, either responses or covariates, as discretised versions of continuous latent variables. We present and compare several models for obtaining these thresholds in the challenging context of count data analysis where the response may be over- and/or under-dispersed in some of the regions of the 
covariate space. We utilise a nonparametric mixture of multivariate Gaussians to model 
the directly observed and the latent continuous variables. 
The paper presents a Markov chain Monte Carlo algorithm for posterior sampling, 
sufficient conditions for weak consistency, and illustrations on density, mean and quantile regression utilizing simulated and real datasets.


\maketitle

\section{Introduction}
We consider methods for Bayesian density regression, with special attention to discrete responses. The basic objective is to estimate the conditional probability mass function of a response $y$ given a vector of covariates $\ux$, $f(y|\ux)$. This allows us to study how the distribution of the response changes with covariates, but of course from the estimated conditional density, 
other quantities of interest, such as the conditional mean, median or other quantiles, may be obtained. 
We approach the problem of conditional density estimation by considering methods for multivariate density 
estimation $f(y,\ux)$ of mixed scale outcomes. From the multivariate density we can obtain the conditional 
density using $f(y|\ux) = f(y,\ux)/\int f(y,\ux)dy$.

Mixture models provide a very popular approach to density estimation. The general form of a mixture 
model for the joint density of $(y,\ux)$ is given by 
\begin{align*}
f_P(y,\ux) = \int_{\uTheta} k(y,\ux;\utheta) dP(\utheta),
\end{align*}
where $k(,;)$ is a probability kernel characterised by parameter $\utheta$ and $P(\cdot)$ is a probability 
measure on the parameter space $\uTheta$. In a Bayesian setting, the model specification is completed by 
specifying a prior distribution for the mixing measure $P(\cdot)$. 

Here we adopt a nonparametric approach in which the prior on $P(\cdot)$ is taken to be a Dirichlet process (DP) 
\citep{Fer73}, resulting in a so-called DP mixture model (DPMM). Due to the discreteness of the DP, 
the DPMM for the joint density can be expressed as
$f_P(y,\ux) = \sum_{h=1}^{\infty} \pi_{h} k(y,\ux;\utheta_h)$. 
DPMMs were introduced by \citet{Lo} and became popular after the work of \citet{EscobarWest} 
and \citet{Mulleretal96}. They allow for flexible modelling of densities without 
having to specify the number of mixture components while at the same time achieving a balance between 
over- and under-fitting.

The implied DPMM for the conditional takes the form of a predictor-dependent mixture 
$f_P(y|\ux) = \sum_{h=1}^{\infty} \pi_{h}(\ux) k(y|\ux;\utheta_h) $, where 
$\pi_{h}(\ux)=\pi_{h} g(\ux;\utheta_h)/f_P(\ux)$, showing how a flexible model for density 
regression is induced by the DPMM for the joint density.  
Furthermore, the mixture formulation allows for complex relationships between $\ux$ and functionals of 
the conditional to be captured. For instance, the implied model for the conditional mean takes the form 
$\text{E}(Y|\ux) = \sum_{h=1}^{\infty} \pi_{h}(\ux) \text{E}(Y|\ux;\utheta_h)$,
which is the approach to nonparametric regression that was introduced by \citet{Mulleretal96}.
Related modelling approaches, which can also be used to analyse discrete data, 
include those of \citet{Shahbaba}, \citet{Taddy}, \citet{DB11} and \citet{Hannah}.

An alternative formulation directly models the conditional $f(y|\ux)$ as a predictor dependent DPMM. 
\citet{DPP} and \citet{CD09} provided such model formulations for continuous responses. Directly modelling 
the conditional distribution can be advantageous over modelling the joint when the covariate vector $\ux$ 
is of high dimension, as the joint model requires estimation of the distribution of the high dimensional vector $(y,\ux^{\top})^{\top}$. A possible remedy to this problem of the joint model is to decompose the overall 
dependence among variables 
into clusters by assuming conditional independence, thereby reducing the problem of estimating a high 
dimensional distribution to the problem of estimating many univariate ones.
On the other hand, the advantages of modelling the joint distribution 
include the simplicity by which a predictor-dependent mixture model is formulated. More importantly, 
modelling the join density provides a direct mechanism for dealing with missing data, as the 
modelled joint $f(y,\ux)$ can be used for imputing missing responses and/or covariates, 
under a missing at random assumption \citep{Kunihama}. Further, the joint formulation,
under a conditional independence assumption, can be used to model a variety of types of covariates, such as 
functional data and images \citep{DB11}. 

Discretization of the latent continuous variables into the observed discrete ones requires the specification of 
either fixed cut-points or of models for the cut-points. There is a growing literature on DPMMs that 
utilise latent variables and fixed cut-points. We note the work of \citet{KMQ2005} and \citet{DeYoreo2} who focus 
on ordinal data, \citet{DeYoreo} who present regression models for binary outcomes, \citet{CanaleDunson14} who 
treat the problem of mixed-scale density estimation from both a theoretical and an applied perspective, and
\citet{NORETS2012332} who also present theory and applications, but based on finite mixture models. However, models 
with fixed cut-points have yet to be extended to include covariates that are fixed by design, such as binary treatment 
allocation variables in clinical trial settings or the offset term in count regression settings. 
Hence, in this paper, we work with modelled cut-points and we introduce 
covariates that are fixed by design through the modelled cut-points, as was 
also done by \citet{PRB}. We consider several models for specifying these cut-points in the context of count data analysis 
and we offer a comparison of their performance in a simulation study. 

Specific models that we consider for the specification of the cut-points, within a  DPMM, include the Poisson, negative binomial 
and generalised Poisson kernels. Whereas the negative binomial can model over-dispersed counts, the generalised Poisson
can model both over- and under-dispersed counts. As was mentioned by \citet{CD11}, DPMMs that utilise the Poisson 
(or negative binomial) kernel, even though they seem very flexible on the surface, in reality are not, 
as they are not be able to adequately model under-dispersed counts. This is our motivation for considering the
three kernels mentioned above. 

Importantly, the paper provides a theoretical examination of the properties of the proposed model by
presenting sufficient conditions for attaining weak consistency. To do so, the paper utilises the theorem of 
\citet[Theorem 6.1]{Schwartz1965} (see also \citet{BNPbook}) and the work of \citet{wu2008}. The conditions that we provide are also sufficient for 
many specials cases of the model presented here, namely DPMM with product mixture kernels that have been utilised 
multiple times in the literature, see e.g. \citet{Taddy} and \citet{DB11}. Furthermore, by utilizing the conditions that we present here as a starting point, one could derive sufficient conditions for weak consistency of DP mixtures of generalised linear models. Surprisingly, this topic that has received very little attention in the literature. To the best of the author's knowledge, the only other paper that considers it is that of \citet{Hannah}.   

The remainder of this paper is arranged as follows. Section \ref{model}
presents the methodology and Section \ref{mcmc} presents a brief
description of the MCMC algorithm we have implemented and methods for posterior inference, 
with most of the details presented in Appendix \ref{mcmcA}. 
Section \ref{wc} presents sufficient conditions for weak consistency, with the proof 
deferred to Appendix \ref{wcp}. 
Sections \ref{simulation} and \ref{application}
present results from a simulation study and an applications to a real dataset. 
The paper concludes with a brief discussion.

\section{Methodology}\label{model}

\subsection{Model specification}

Let $Y$ denote a discrete response with support on (a subset of) the non-negative integers
and let $\uX=(\uX_{d}^{\top},\uX_{c}^{\top})^{\top}$ denote a vector of $p$ mixed scale covariates,
where $\uX_{d}$ is a vector of $p_d$ discrete variables and $\uX_{c}$ is a vector of $p_c$ continuous
variables, $p_d+p_c = p$.  
Our goal is to jointly model $\uZ = (Y,\uX^{\top})^{\top}$ as a draw from an 
unknown density $f \in \mathcal{F}$ with respect to an appropriate measure, where $\mathcal{F}$
denotes the set of all such densities. The next few paragraphs describe how a prior $\Pi$
on $\mathcal{F}$ is induced. 

Discrete variables, either responses or covariates, are assumed to be 
discretised or rounded versions of continuous latent variables \citep{Muthen}.  
Our presentation below concerns generic discrete and continuous latent variables
denoted by $Z$ and $Z^*$ respectively. 
Observed and latent variables are connected according to the rule
\begin{align*}
Z = z \text{\;\;if and only if\;\;} z^* \in R(z) = (c_{z-1}, c_{z}], z=0,1,2,\ldots,
\end{align*}
where $R(z)$ is an interval on the real line with bounds given by: 
$c_{-1} = -\infty$, and for $q \geq 0$, $c_{q} = c_{q}(\ulambda) = \Phi^{-1}\{F(q;\ulambda)\}.$ 
Here $\Phi(\cdot)$ is the cumulative distribution function of a standard normal variable, and $F(\cdot;\ulambda)$ 
denotes an appropriate cumulative distribution function. 
Further, latent variables are assumed to be independent draws from
a N$(0,1)$ distribution, where the mean and variance are restricted to be zero and one respectively
as they are non-identifiable by the data. It is easy to see that with this specification the marginal distribution 
of $Z$ is $F(z;\ulambda)$:
\begin{align}
\Pr(Z \leq z) = \Pr\{Z^* < \Phi^{-1}[F(z;\ulambda)]\} = \Pr[\Phi(Z^*) < F(z;\ulambda)]
= F(z;\ulambda),\label{marginal}
\end{align}
where the last equality follows because $\Phi(Z^*)$ has a uniform distribution on the unit interval
(see e.g. \citet{RC05}). 

Generally, equation (\ref{marginal}) is satisfied if one assumes $c_{q} = G^{-1}[F(q;\ulambda)]$ and that the latent variables are 
independent draws from $G$, where $G$ is a continuous cumulative distribution function. Common choices for the density of the latent variables $Z^*$ include the Student's t, Weibull, lognormal and gamma. In this paper, we focus on the case where $G=\Phi$ and we examine alternatives for function $F(\cdot;\ulambda)$. The next few paragraphs discuss choices for $F(\cdot;\ulambda)$ for count and binomial data. 

\emph{Count data:} For modelling counts, we may take $F(\cdot;\ulambda)$ to be the distribution function of a Poisson$(H \xi_1)$ variable. 
Here $\ulambda=(\xi_1,H)^{\top}$, where $\xi_1 > 0$ denotes the Poisson rate and $H$ the offset term. Hence, the associated probability mass function (pmf) is given by 
$\Pr(Z=z;\xi_1) = \exp(-H\xi_1) (H\xi_1)^z /z!$, with implied mean and variance both equal to $H\xi_1$. 
Further, to account for potential over-dispersion in the counts, we may take $F(\cdot;\ulambda)$ to be the negative binomial distribution function. With this choice, the 
vector $\ulambda=(\xi_1,\xi_2,H)^{\top}$ includes the offset term $H$ and two unknown parameters $\uxi=(\xi_1,\xi_2)^{\top}$ that 
allow for extra flexibility compared to the flexibility afforded by the single parameter Poisson distribution. 
The pmf is given by 
\begin{align}
\Pr(Z=z;\xi_1,\xi_2) = \frac{\Gamma(z+\xi_1)}{\Gamma(\xi_1)\Gamma(z+1)} \left(\frac{\xi_2}{H+\xi_2}\right)^{\xi_1} \left(\frac{H}{H+\xi_2}\right)^{z}, \nonumber
\end{align}
where $\xi_1>0$ and $\xi_2 > 0$, and it implies a mean and a variance of $H \xi_1/\xi_2$ and $H \xi_1/\xi_2 (1 + H/\xi_2)$, respectively. 
Both of these choices, however, are quite restrictive as they require the variance to be equal or greater than the mean. 
Under-dispersion cannot be modelled in a satisfactory way even with nonparametric mixtures of Poisson or negative binomial pmfs. 
For this reason, we also consider the generalised Poisson \citep{CF92} distribution function  
that allows for both over- and under-dispersion relative to the Poisson. The pmf is given by 
\begin{align}
\Pr(Z=z;\xi_1,\xi_2) = \xi_1 [\xi_1+(\xi_2-1)z]^{z-1} \xi_2^{-z} \exp\{-[\xi_1+(\xi_2-1)z]/\xi_2\}/z!,\label{gp}
\end{align}
where $\xi_1>0$ and $\xi_2 > 0$. It may be shown that the pmf implies that $\text{E}(Z) = \xi_1$ and 
var$(Z) = \xi_2^2 \xi_1$. Hence, the distribution is over-dispersed when $\xi_2 > 1$, under-dispersed when $\xi_2 < 1$ 
and it reduces to the Poisson pmf when $\xi_2=1$. 
When $\xi_2 < 1$ an upper bound $m$ is set on the counts, where $m$ is the largest integer for which
$\xi_1+(\xi_2-1)m > 0$, so that the pmf remains positive on its support. In this case the normalizing constant of the pmf in (\ref{gp}) 
needs to be computed because the sum of probabilities $\sum_{z=0}^m \Pr(z;\xi_1,\xi_2)$ is not necessarily 
equal to one. Lastly, it is straight forward to include an offset term $H$ in this pmf, by replacing $\xi_1$ by $H\xi_1$.  

\emph{Binomial data:} For modelling binomial data, we may take $F(\cdot;\ulambda)$ to be the distribution function of a 
Binomial$(H,\xi_1)$ variable, with $\xi_1$ denoting the success probability and $H$ the number of trials. 
A more flexible approach would be to take $F(\cdot;\ulambda)$ to be the beta-binomial distribution function, where $\ulambda=(\xi_1,\xi_2,H)^{\top}$
includes two unknown parameters. The associated pmf is given by
\begin{align*}
\Pr(Z=z;\xi_1,\xi_2) = \binom{H}{z} \frac{\text{Beta}{(z+\xi_1,H-z+\xi_2)}}{\text{Beta}{(\xi_1,\xi_2)}}, 
\end{align*}
where $\xi_1>0$ and $\xi_2 > 0$.

\emph{Binary data:} The special case of binomial data with $H=1$ can equivalently be treated as 
\begin{equation}
\begin{split}
&Z = 0 \text{\;\;if and only if\;\;} z^* < 0, \\
&z^* \sim \text{N}(\mu_z^{*},1), \label{binary}
\end{split}
\end{equation}
which is the approach of our preference as it allows for simpler posterior sampling. 

We induce a prior $\Pi$ on $\mathcal{F}$ by assuming  a nonparametric mixture model 
for $\uZ = (Y, \uX_{d}^{\top}, \uX_{c}^{\top})^{\top},$ 
\begin{align}
f_P(\uz) = \int_{\uTheta} k(\uz;\utheta) dP(\utheta) \label{one}
\end{align} 
that utilises a parametric kernel $k(\uz;\utheta)$ and a nonparametric model for the random mixing 
distribution $P$ on $\uTheta$. 

The kernel $k(\uz;\utheta)$ is obtained by assuming a $q$-dimensional Gaussian for the continuous observed 
and latent variables, $\uZ^* = (Y^*, \uX_{d}^*{^{\top}}, \uX_{c}^{\top})^{\top},$  and integrating out the latent variables
\begin{align}\label{kernel3}
k(\uz;\utheta) = \int_{R(y)} \int_{R(x_{d})} 
\text{N}(\uz^*;\umu^*,\uSigma^*) d\ux_{d}^{*} dy^{*},
\end{align} 
where $q=1+p$ and $\utheta = (\uxi,\umu^*,\uSigma^*)$ denotes the kernel parameters.
Due to the non-identifiability of the location and scale parameters of the distribution of 
the latent variables, with the exception of the location parameter in (\ref{binary}), 
the mean $\umu^*$ and covariance $\uSigma^*$ are of the form
\begin{align}\label{restrictedMCV}
\begin{array}{ll}
\umu^*=\left( 
\begin{array}{l}
\uzero \\
\umu \\
\end{array}
\right),
&
\uSigma^*=\left[ 
\begin{array}{ll}
\uC &  \unu^{\top} \\
\unu & \uSigma \\
\end{array}
\right],
\end{array}
\end{align}
where $\uC$ is the covariance matrix of the latent continuous variables and has  
diagonal elements equal to one i.e. it is a correlation matrix.
Further, $\uSigma$ is the unrestricted covariance matrix of the directly observed continuous variables.
Specific examples are provided later in the paper and they concern: (i) a count response and a
continuous covariate--see (\ref{exmpl1}); and (ii) a count response and a binary and a continuous covariate--see (\ref{exmpl2}). Below we consider some special cases, where the kernel can be simplified, and draw connections
to the literature.

As a first special case we consider the so-called product kernel, obtained when $\unu=\uzero$ and 
$\uC=\uI$, where $\uI$ is the identity matrix. The choice $\uC=\uI$ implies that the discrete variables 
are conditionally independent and $\unu=\uzero$ implies that discrete and continuous variables are 
conditionally independent. Further, assuming that $\uSigma$ is diagonal implies that the continuous variables 
are conditionally independent. Within a Bayesian nonparametric framework, such kernels have been utilised 
by \citet{Taddy} and \citet{DB11}.

As another special case we consider the scenario where all $p$ covariates are continuous. 
Here the joint mean is $\umu^*=(\mu_y,\umu_x^{\top})^{\top}$ and the joint covariance $\uSigma^*$
has submatrix $\uC=1$, $\unu$ a $p$-dimensional vector, and $\uSigma$ a $p \times p$ positive definite matrix. 
The kernel in (\ref{kernel3}) may be written as 
\begin{align}\label{kernel2}
k(\uz;\utheta) = \int_{R(y)} \text{N}(\uz^*;\umu^*,\uSigma^*) dy^{*} =
\text{N}(\ux;\umu,\uSigma) \int_{R(y)}  \text{N}(y^{*};m^*,v^*) dy^{*}, 
\end{align} 
where 
$m^* = \text{E}(y^{*}|\ux)=\mu_y + \unu^{\top} \uSigma^{-1} (\ux-\umu)$ and
$v^* = \text{var}(y^*|\ux)=1-\unu^{\top} \uSigma^{-1}\unu$.

When the response is binary, (\ref{kernel2}) becomes
\begin{align*}
k(Y=1,\ux;\utheta) = \text{N}(\ux;\umu,\uSigma) \Phi(m^*/\sqrt{v^*}),
\end{align*} 
implying a probit regression model for the conditional probability of success.
A Bayesian non-parametric approach for binary regression based on this kernel
has been developed by \citet{DeYoreo}.

\subsection{Prior specification}

Following a Bayesian nonparametric approach, we assign to the unknown mixing distribution $P(\cdot)$ a 
Dirichlet process (DP) prior \citep{Fer73}. A DP prior is characterised by two parameters: a total mass 
or concentration parameter $\alpha$ and a base distribution $P_0$ over the parameter space. 
According to the stick-breaking representation \citep{sethuraman} 
\begin{align*}
P(\cdot)=\sum_{h=1}^\infty \pi_{h} \delta_{\utheta_h}(\cdot),
\end{align*}
which when combined with (\ref{one}) leads to the following DPMM 
for $\uz=(y,\ux_{d}^{\top},\ux_{c}^{\top})^{\top}$ 
\begin{align}
f_P(\uz) = \sum_{h=1}^{\infty} \pi_{h} k(\uz;\utheta_h). \label{slice1}
\end{align}
In the above countable mixture, the weights $\pi_{h}, h \geq 1,$ are constructed by the so-called stick-breaking process:
$\pi_1 = v_1$, and for $l \geq 2$, $\pi_l = v_l \prod_{h=1}^{l-1} (1-v_h)$,
where $v_k, k \geq 1$, are independent draws from a $\text{Beta}(1,\alpha)$ distribution. Further,  
the atoms $\utheta_h, h \geq 1,$ are obtained as independent draws from the base distribution $P_0$, which consists of 
three independent priors for the elements of $\utheta_h = (\uxi_h, \umu_h, \uSigma^*_h),$ $h \geq 1$,
described next.

Firstly, the priors on the set of parameters $\{\uxi_h\}$ depend on the choice of the function $F(;)$ in (\ref{marginal}). 
For all functions, we take these priors to be very close to uninformative. Table \ref{fam2} provides a summary.  
For the rate $\xi_1$ of the Poisson distribution, we take the prior to be Gamma$(1,0.1)$, a gamma distribution with mean $10$ and variance $100$. 
The same prior is taken for the $\xi_1$ and $\xi_2$ parameters of the negative binomial distribution function. Further, the mean parameter of the generalised Poisson
distribution, $\xi_1$, is given the same gamma prior, while the dispersion parameter, $\xi_2$, is given a normal prior with mean and variance equal to
$1$, and truncated from below at $0.05$. In addition, the binomial probability of success is given a uniform prior, while the two parameters of the beta-binomial
distribution are given vague gamma priors.  
\begin{table}[htp]
	\caption{\label{fam2} 
		Prior distributions on the pmf parameters.} 
	\centering		
		\begin{tabular}{cll}
			& pmf & Prior distribution\\
			\midrule
			1. & Poisson$(\xi_1)$ & $\xi_1 \sim$ Gamma$(1,0.1)$\\
			2. & Negative binomial$(\xi_1,\xi_2)$ & $\xi_1, \xi_2 \sim$ Gamma$(1,0.1)$\\
			3. & Generalised Poisson$(\xi_1,\xi_2)$ & $\xi_1 \sim \text{Gamma}(1,0.1)$\\
			&                                 & $\xi_2 \sim \text{N}(1,1) \uone(\xi_2 > 0.05)$\\
			4. & Binomial$(\xi_1)$ & $\xi_1 \sim$ Beta$(1,1)$\\
			5. & Beta-binomial$(\xi_1,\xi_2)$ & $\xi_1, \xi_2 \sim$ Gamma$(1,0.1)$\\
			\bottomrule
		\end{tabular}
\end{table}

Secondly, the prior on $\umu_h$, the non-zero part of $\umu_h^*$, is taken to be multivariate normal $\umu_h \sim \text{N}(\ud,\uD)$. 
The mean $\ud$ is taken to be equal to the centre of the dataset. Specifically, the part of $\ud$ that corresponds to 
continuous variables is taken to be equal to the sample mean while the part that corresponds to binary variables is taken to be 
a transformation of the observed sample proportion. Let $p^*$ denote an observed sample proportion. The corresponding prior mean
is taken to be $-\Phi^{-1}(1-p^*)$, which along with the mechanism in (\ref{binary}) implies a prior proportion equal to 
$p^*$. Further, the covariance matrix $\uD$ is taken to be diagonal. Its elements that correspond to variances of continuous variables are set equal to a small multiple (here taken to be $1/8$) of the square of the observed data range \citep{RG1997},  
while the elements that correspond to binary variables are set equal to a constant (here taken to be $5$). 

Lastly, the prior distribution assigned to the restricted covariance matrices $\uSigma_{h}^{*}, h \geq 1,$
in (\ref{restrictedMCV}) is specified by utilizing the methods of \citet{xiao} and \citet{Barnard00}: 
we add into the model variance parameters that are non-identifiable by the data 
and then separate identifiable from non-identifiable parameters. 
The starting point is a Wishart$(\uE_h;\eta,\uEta)$ prior for unrestricted $q \times q$ 
covariance matrices $\uE_h, h \geq 1$:
\begin{align*}
p(\uE_h;\eta,\uEta) \propto |\uE_h|^{(\eta-q-1)/2} 
\exp[\tr(-\uEta^{-1} \uE_h/2)],
\end{align*}
where 
\begin{align*}
\uH =
\left[ 
\begin{array}{ll}
\uH_{11} & \uH_{12} \\
\uH_{12}^{\top} & \uH_{22} \\
\end{array}
\right],
\end{align*}
where $\uH_{11}$ is a $(1+p_d) \times (1+p_d)$ correlation matrix, $\uH_{22}$ is a $p_c \times p_c$ 
unrestricted covariance matrix, and $\uH_{12}$ is a $(1+p_d) \times p_c$ matrix of covariances.

We decompose $\uE_h=\uD_h^{1/2} \uSigma_h^{*} \uD_h^{1/2}$, where 
$\uD_h = \text{Diag}(d^2_{h,1},\ldots,d^2_{h,1+p_d},1,\ldots,1)$ is 
a diagonal matrix of $(1+p_d)$ non identifiable variance parameters and $p_c$ ones that
correspond to identifiable variances, and $\uSigma_h^{*}$ is 
a covariance matrix that satisfies the restrictions imposed by the data.
The Jacobian of this transformation is 
$J(\uE_h \rightarrow \uD_h, \uSigma_{h}^{*}) = \prod_{j=1}^{(1+p_d)} d_{hj}^{q-1} = |\uD_h|^{(q-1)/2}$. 
Hence, we obtain the following prior for the pairs
$(\uD_h, \uSigma_{h}^{*}), h \geq 1$: 
\begin{align*}
p(\uD_h, \uSigma_{h}^{*};\eta,\uEta) \propto |\uE_h|^{(\eta-q-1)/2} 
\exp[\tr(-\uEta^{-1} \uE_{h}/2)] J(\uE_{h} \rightarrow \uD_h, \uSigma_{h}^{*}).
\end{align*}
In our analyses we take $\eta$ to be equal to $q+2$ and $\uEta$ to be diagonal, with sub-matrix $\uH_{11}$ equal
to the identity matrix and with sub-matrix $\uH_{22}$ having entries equal to a small multiple ($1/8$) of the square of 
the observed data range.  

\section{Posterior sampling and inference}\label{mcmc}

\subsection{MCMC sampler}

The main tools that we utilise in developing an MCMC sampler are the `blocked' approach of  
\citet{IJ01} and adaptive Metropolis algorithms \citep{roberts_examples_2009} to achieve
optimal scaling \citep{Roberts2001c}.

We start by truncating the countable mixture in (\ref{slice1}) to include $T$ components 
and by introducing an allocation variable $\delta$. The model is now written in the following 
equivalent way
\begin{align*}
&\uz | \utheta, \delta=l \sim k(\uz;\utheta_{l}), \\ 
&\Pr(\delta = l|\alpha) =\pi_{l}, l = 1,2,\ldots,T. \end{align*}
The likelihood associated with $n$ independent and identically distributed observations $\uz_i, i=1,2,\ldots,n,$ 
can be written as
\begin{align*}
\ell(\utheta,\alpha;\uz,\udelta) = \ell(\utheta,\alpha;\uz_{i},\delta_{i}=l_{i}, i=1,\ldots,n) = 
\prod_{i} k(\uz_{i};\utheta_{l_{i}}) \pi_{l_{i}}.
\end{align*}
Recall that $\uz_i$ consist of a discrete response $y_i$, $p_d$ discrete covariates 
$\ux_{d,i} = (x_{d,i,1},\ldots,x_{d,i,p_d})^{\top}$ and $p_c$ continuous covariates
$\ux_{c,i} = (x_{c,i,1},\ldots,x_{c,i,p_c})^{\top}, i=1,2,\ldots,n$. Augmenting the likelihood 
with the latent continuous variables $y_i^*$ and $\ux_{d,i}^*= (x_{d,i,1}^*,\ldots,x_{d,i,p_d}^*)^{\top}$, we obtain
\begin{align*}
&\ell(\utheta,\alpha;\uz,\udelta,\uy^*,\ux_d^*) = \\
&\prod_{i} \left\{\uone(y_{i}^{*} \in R(y_i)) \left[\prod_{m=1}^{p_d} \uone(x_{d,i,m}^{*} \in R(x_{d,i,m})) \right]
\text{N}(y_{i}^{*},\ux_{d,i}^{*},\ux_{c,i};\umu^*_{l_{i}},\uSigma^*_{l_{i}})\pi_{l_{i}}\right\},
\end{align*}
from which the full posterior follows 
\begin{align*}
\pi(\utheta,\udelta,\uy^*,\ux_d^*,\alpha|\uy,\ux_d,\ux_c) \propto \ell(\utheta,\alpha;\uz,\udelta,\uy^*,\ux_d^*) p_0(\utheta,\alpha).
\end{align*}

\subsection{An exact algorithm}

Truncation of the mixture density in (\ref{slice1}) can be avoided by implementing a slice sampler
\citep{Walker07, Omiros}. This requires augmenting the likelihood with uniform random variables 
$U_i \sim \text{U}(0,1) , i=1,\ldots,n,$ such that the complete likelihood becomes
\begin{align*}
\prod_{i} \left\{\uone(y_{i}^{*} \in R(y_i)) \left[\prod_{m=1}^{p_d} \uone(x_{d,i,m}^{*} \in R(x_{d,i,m})) \right]
\text{N}(y_{i}^{*},\ux_{d,i}^{*},\ux_{c,i};\umu^*_{l_{i}},\uSigma^*_{l_{i}}) \uone(u_i < \pi_{l_{i}})
\right\}.
\end{align*}
Details on the updating steps of the MCMC algorithm are provided in Appendix \ref{mcmcA}.
We note that throughout the paper we utilise the truncated sampler as truncation allows  
sampling from the full posterior distribution, including sampling from the posterior 
of the random mixing distribution $P$, which in turn allows for proper uncertainty quantification for density estimates. The same strategy was utilised by \citet{DeYoreo2}.

\subsection{Posterior inference}
Recall that $\uz = (y,\ux_d^{\top},\ux_c^{\top})^{\top}$ and $\uz^* = (y^*,\ux_d^{*^{\top}},\ux_c^{\top})^{\top}$.
Then the model for $\uz$, truncated to include $T$ components, is expressed as 
\begin{align}
f_P(\uz) = \sum_{h=1}^{T}\pi_h k(\uz;\utheta_h) = \sum_{h=1}^{T}\pi_h \int_{R(y)} \int_{R(x_d)} \text{N}(\uz^*;\umu_h^*,\uSigma_h^*)dy^* d\ux_d^*.\label{joint}
\end{align}
Further, let $\umu_h^*$ and $\uSigma_h^*$ be partitioned as follows 
\begin{align*}
\begin{array}{ll}
\umu_h^*=\left( 
\begin{array}{c}
\umu_{h,d} \\
\umu_{h,x_c} \\
\end{array}
\right),
&
\uSigma_h^*=\left[ 
\begin{array}{ll}
\uSigma_{h,d,d} &  \uSigma_{h,d,x_c} \\
\uSigma_{h,x_c,d} & \uSigma_{h,x_cx_c}  \\   
\end{array}
\right],
\end{array}
\end{align*}
where subscript $d$ denotes the continuous latent variables underlying discrete ones $(y^*,\ux_d^{*^{\top}})^{\top}$. 
Then (\ref{joint}) may be expressed as 
\begin{align}
\sum_{h=1}^{T} \pi_h \text{N}(\ux_c;\umu_{h,x_c},\uSigma_{h,x_cx_c}) \int_{R(y)} \int_{R(x_d)} \text{N}(y^*,\ux_d^*|\ux_c;\umu_{h,d.c},\uSigma_{h,d.c})dy^*d\ux^*_d,\label{num1}
\end{align}
where $\umu_{h,d.c} = \umu_{h,d} + \uSigma_{h,d,x_c}  \uSigma_{h,x_cx_c}^{-1} (\ux_c - \umu_{h,x_c})$ and 
$\uSigma_{h,d.c} = \uSigma_{h,d,d} - \uSigma_{h,d,x_c} \uSigma_{h,x_cx_c} ^{-1} \uSigma_{h,x_c,d}$.

Further, to obtain an expression for the conditional $f_P(y|\ux)$, first let $\ux = (\ux_d^{\top},\ux_c^{\top})^{\top}$ and $\ux^* = (\ux_d^{*^{\top}},\ux_c^{\top})^{\top}$.
In addition, let $\umu_{h,x}^*$ and $\uSigma_{h,xx}^*$ denote the mean and covariance of $\ux^*$, which we partition as
\begin{align*}
\begin{array}{ll}
\umu_{h,x}^*=\left( 
\begin{array}{c}
\umu_{h,x_d} \\
\umu_{h,x_c} \\
\end{array}
\right),
&
\uSigma_{h,xx}^*=\left[ 
\begin{array}{ll}
\uSigma_{h,x_dx_d} &  \uSigma_{h,x_dx_c} \\
\uSigma_{h,x_cx_d} & \uSigma_{h,x_cx_c}  \\   
\end{array}
\right].
\end{array}
\end{align*}

Now, the conditional $f_P(y|\ux)$ may be expressed as  
\begin{align*}
f_P(y|\ux) = \frac{f_P(\uz)}{f_P(\ux)} = \frac{\sum_{h=1}^{T}\pi_h k(\uz;\utheta_h)}{\sum_{h=1}^{T}\pi_h k(\ux;\utheta_{h})},
\end{align*}
where $k(\ux;\utheta_{h}) = \int_{R(x_d)} \text{N}(\ux^*;\umu_{h,x}^*,\uSigma_{h,xx}^*)d\ux^*_d$. Lastly, utilizing 
a similar factorization as in (\ref{num1}),
we may write
\begin{align*}
k(\ux;\utheta_{h}) = \text{N}(\ux_c;\umu_{h,x_c},\uSigma_{h,x_cx_c}) \int_{R(x_d)} \text{N}(\ux_d^*|\ux_c;\umu_{h,x_{d.c}},\uSigma_{h,x_{d.c}})d\ux^*_d,
\end{align*}
where $\umu_{h,x_{d.c}} = \umu_{h,x_d} + \uSigma_{h,x_dx_c}  \uSigma_{h,x_cx_c}^{-1} (\ux_c - \umu_{h,x_c})$ and 
$\uSigma_{h,x_{d.c}} = \uSigma_{h,x_dx_d} - \uSigma_{h,x_dx_c} \uSigma_{h,x_cx_c} ^{-1} \uSigma_{h,x_cx_d}$. 
Hence, we find that $f_P(y|\ux)$ can be expressed as 
\begin{align*}
\frac{\sum_{h=1}^{T} \pi_h \text{N}(\ux_c;\umu_{h,x_c},\uSigma_{h,x_cx_c}) \int_{R(y)} \int_{R(x_d)} \text{N}(y^*,\ux_d^*|\ux_c;\umu_{h,d.c},\uSigma_{h,d.c})dy^*d\ux^*_d}
{\sum_{h=1}^{T} \pi_h \text{N}(\ux_c;\umu_{h,x_c},\uSigma_{h,x_cx_c}) \int_{R(x_d)} \text{N}(\ux_d^*|\ux_c;\umu_{h,x_{d.c}},\uSigma_{h,x_{d.c}})d\ux^*_d}.
\end{align*}

Of interest is the quantity
\begin{align*}
f(y|\ux) = \int \frac{f_P(\uz)}{f_P(\ux)} dP.   
\end{align*}
Given samples from the posterior of $P$, denoted by $P_s, s=1,\ldots,S,$ $f(y|\ux)$ will be approximated by 
\begin{align*}
f(y|\ux) = S^{-1} \sum_{s=1}^S f_{P_s}(y|\ux).
\end{align*} 

For each sampled conditional $f_{P_s}(y|\ux)$ we calculate all functionals 
of interest and thereby obtain posteriors for these functionals. This process is carried out on a grid of $\ux$ values,
enabling inference about the dependence of the conditional $f(y|\ux)$ and its functionals on $\ux$.

\section{Weak Consistency}\label{wc}

Here we provide sufficient conditions under which the proposed mixture model for the joint density attains weak 
posterior consistency at the true distribution $f_0$. Weak consistency refers to the property of the posterior distribution
to concentrate in regions of $\mathcal{F}$ that are close to the true distribution $f_0$ in the weak topology sense.
To formalise this concept we next provide some basic definitions, followed by two important theorems based on which
we establish weak consistency for the proposed model.  

Recall that $\mathcal{F}$ denotes the space of mixed scale densities with respect to some suitable measure. 
A weak neighbourhood of $f_0 \in \mathcal{F}$ of radius $\epsilon$ is defined as
\begin{align*}
V_{\epsilon}(f_0) = \left\{f \in \mathcal{F}: \left|\int \phi_i f - \int \phi_i f_0 \right| < \epsilon, i=1,\ldots,m \right\},
\end{align*}
where $\phi_i, i=1,\ldots,m,$ are bounded, continuous functions. 

Further recall that $\Pi$ is a prior on $\mathcal{F}$ and let $\uZ_1,\ldots,\uZ_n$ be i.i.d. with common density $f_0$.
The posterior probability of $A \subset \mathcal{F}$ is given by 
\begin{align*}
\Pi(A|\uZ_1,\ldots,\uZ_n) = \frac{\int_A \prod_{i=1}^n f(\uZ_i)\Pi(d f)}{\int_{\mathcal{F}} \prod_{i=1}^n f(\uZ_i)\Pi(d f)}.
\end{align*}

A prior $\Pi$ is said to be weakly consistent at $f_0$ if
\begin{align*}
\Pi(V_{\epsilon}|\uZ_1,\ldots,\uZ_n) \rightarrow 1
\end{align*}
for all weak neighbourhoods $V_{\epsilon}$ of $f_0$, with $\mathcal{P}_{f_0}$ probability 1. 

The Kullback-Leibler (KL) neighbourhood of $f_0 \in \mathcal{F}$ of radius $\epsilon$ is defined as follows
\begin{align*} 
K_{\epsilon}(f_0) = \left\{f \in \mathcal{F}: \int f_0 \log(f_0/f) < \epsilon \right\}.
\end{align*}
A density $f_0$ is said to be in the KL support of $\Pi$ if $\Pi(K_{\epsilon}(f_0))>0$ for all
$\epsilon > 0$.

To establish weak consistency we utilise the following theorem of \citet{Schwartz1965}.

\textbf{Theorem}: If $f_0$ is in the KL support of $\Pi$, then the posterior is weakly consistent at $f_0$.

To prove the KL property for the mixture model described in this paper, we utilise Theorem 1 of \citet{wu2008}, stated below. 

\textbf{Theorem}: Let $f_0(y,\ux)$ denote the true density, $\Pi^*$ the prior on $\mathcal{M}(\uTheta)$, the space of probability measures on $\uTheta$,
and $\Pi$ the prior induced on $\mathcal{F}$. If for any $\epsilon > 0$, there exists mixing distribution $P_{\epsilon}$ and a
$\mathcal{W} \subset \mathcal{M}(\uTheta)$ with $\Pi^*(\mathcal{W})>0$, such that 
\begin{enumerate}
	\item[$A_1$:] $\;\sum_y \int f_0(y,\ux) \log[f_0(y,\ux)/f_{P_{\epsilon}}(y,\ux)]d\ux < \epsilon$,
	\item[$A_2$:] $\;\sum_y \int f_0(y,\ux) \log[f_{P_{\epsilon}}(y,\ux)/f_P(y,\ux)]d\ux < \epsilon$ for every $P \in \mathcal{W}$,
\end{enumerate}
then $f_0 \in \text{KL}(\Pi)$.   

The main result on the weak consistency of the proposed model is stated in the following lemma.
It is based on a special case of the overall model, namely the case where there is a discrete response and 
$p$ continuous covariates, and hence the kernel is the one that appears in (\ref{kernel2}).  

\textbf{Lemma}: Under the following conditions we may establish that the prior defined in (\ref{one}) and (\ref{kernel3}) 
satisfies conditions $A_1$ and $A_2$ of the theorem of \citet{wu2008}:
\begin{enumerate}
	\item[$C_1$:] \; the true density $f_0(y,\ux)$ can be expressed as $f_0(y,\ux) = \int K(y;\uxi) f_0^*(\uxi,\ux)d\uxi$ and 
	it is compactly supported,
	\item[$C_2$:] \; the density $f_0^*$ is continuous, compactly supported and it satisfies $0<f_0^*(\uz)<M$ almost everywhere,
	\item[$C_3$:] \; there exists an $m$ such that $\inf_{||\ux||\leq m} f_0^*(\ux|\uxi) \geq c$ for all $\uxi$,
	\item[$C_4$:] \; for $\ux$ such that $||\ux||>m$, $f^*_0(\uxi,\ux)<c$. Further, $f_0^*$ is decreasing as $||\ux||$ 
	increases more than $m$,
	\item[$C_5$:] \; $|\sum_y \int_{x} f_0(y,\ux)\log f_0(y,\ux)d\ux| < \infty$.
\end{enumerate}
Hence, $f_0 \in \text{KL}(\Pi)$ and by the theorem of \citet{Schwartz1965}, the posterior is weakly consistent at $f_0$.
The proof, which is provided in Appendix \ref{wcp}, clarifies why these five conditions are needed. 

\section{Simulation study}\label{simulation}

Here we present results from a simulation study. The key aim is to provide insights into the effect of 
the kernel choice on posterior inference for a regression surface and other functionals of the conditional pmfs.

\begin{figure}
	\begin{center}
		\includegraphics[width=0.45\textwidth]{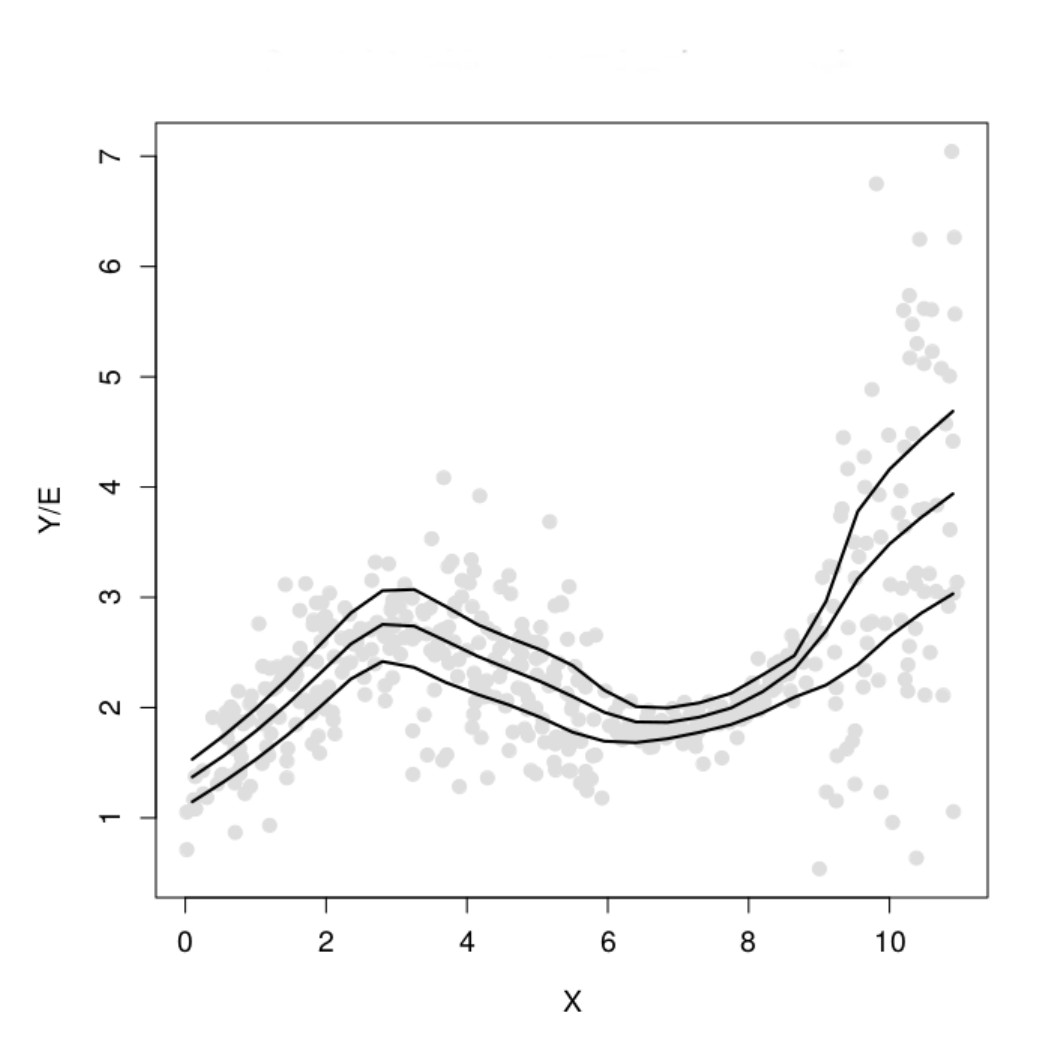}\\
	\end{center}
	\caption{Simulation study results obtained from the DPMM with the generalised Poisson distribution function.
		The three curves show the results concerning the mean and $25$th and $75$th percentiles.}
	\label{figure1.sim}
\end{figure}

The data-generating mechanism consists of a continuous covariate $X \sim \text{U}(0,11)$ and a count
response $Y$, which has the following conditional mean function
\begin{align*}
\mu_x = \text{E}(Y|X=x) = 1 + \sin(\pi x/5) + x/4.
\end{align*}
The mechanism from which we generate the responses is
\begin{align*}
Y | X=x \sim \left\{ 
\begin{array}{ll}
\text{Poisson}(H \mu_x), & x < 3, \\
\text{Poisson}(H \mu_x |\epsilon_1|), & 3 < x < 6, \\
\text{round}(H \mu_x + \epsilon_2), & 6 < x < 9, \\
\text{Poisson}(H \mu_x |\epsilon_3|), & x > 9,\\
\end{array}
\right.
\end{align*}
where $H$ denotes the offset term, here generated from $H \sim \text{U}(10,30)$. 
Furthermore, round$(\cdot)$ is a function that rounds its argument to the closest integer and  
$\epsilon_i, i=1,2,3,$ are normally distributed random errors: $\epsilon_1 \sim \text{N}(1,0.15^2)$, 
$\epsilon_2 \sim \text{N}(0,2^2,)$ and $\epsilon_3 \sim \text{N}(1,0.30^2)$.
Because the mechanism places positive probability on negative realizations $y$,
we take the realised response to be $\max(0,y)$. The mechanism generates responses that are Poisson distributed over the range $x<3$,
`mildly' over-dispersed over $3<x<6$, under-dispersed over $6<x<9$, and `severely' over-dispersed relative to the Poisson distribution
over $x>9$. We take the sample size to be $n=500$. A simulated dataset is shown in Figure \ref{figure1.sim}.  

For each simulated dataset we fit a model of the form 
\begin{align}
&&f(y^*_i,x_{i}) = \sum_{h=1}^{\infty} \pi_h 
\text{N}\left(
\begin{array}{ll}
\left[ 
\begin{array}{l}
0 \\
\mu_{h} 
\end{array}
\right],
&
\left[ 
\begin{array}{ll}
1.0 &  \sigma_{12} \\
\sigma_{21} &  \sigma_{22}\\
\end{array}
\right]
\end{array}\right)\label{exmpl1},
\end{align}
where $y^*_i$ denotes the latent continuous variable underlying the count response $y_i, i=1,\ldots,n$. 
Observed and latent responses are connected by 
$Y_i = y_i$ if and only if $c_{y_i-1} < y_i^* \leq c_{y_i}, \text{where\;} c_{y_i} = \Phi^{-1}[F(y_i;H,\uxi)]$. 
We consider three choices for the function $F(;,)$ that appears in the definition of the cut-points. These
are the Poisson, negative binomial (NB) and generalised Poisson (GP) distribution functions. 
We note that the data generating mechanism is not nested within any of the models we fit.

Results presented are based on $20$ replicate datasets. For each dataset and for each choice of 
$F(;,)$ we obtained $40,000$ posterior samples of which we discarded the first $20,000$ as burn-in. 
Of the remaining $20,000$ samples, we retained one every $5$. Furthermore, during posterior simulation 
we obtained samples of conditional pmfs $f(Y|X=x)$ for $25$ values of $x$ equally spaced between $0.1$ 
and $10.90$ and for offset term $H$ equal to 
the mean of the sampled offset terms. For each sampled conditional pmf, we calculated the mean and $25$th and $75$th percentiles. 
We compared sampled and true values of these functionals by calculating the medians of the posteriors of the total squared errors  
$\{\sum_{c=1}^{25} (p_c-p_c^{rs})^2: s=1,\ldots,4000, r=1,\ldots,20\}$, where $p_c$ denotes the true value of the parameter of interest, 
here the mean, $25$th and $75$th percentile, of the $c$th conditional 
pmf, $c=1,\ldots,25,$ and $p_c^{rs}$ denotes the $s$th sampled value of the parameter of interest,  
$s=1,\ldots,4000,$ when fitting the model to the $r$th replicate dataset, $r=1,\ldots,20.$ 

\begin{table}[htp]
		\caption{\label{table1.sim}
			Simulation study results: posterior medians of total squared errors for the three functionals of interest and the three kernels.} 
		\centering
		\begin{tabular}{l r@{.}l r@{.}l r@{.}l }
			& \multicolumn{2}{c}{Mean}  & \multicolumn{2}{c}{Q$_{25}$} & \multicolumn{2}{c}{Q$_{75}$}\\
			\midrule                
			Poisson              &  0&762 & 0&796 & 1&587\\ 
			Negative binomial     &  0&513 & 0&862 & 1&428\\  
			Generalised Poisson  &  0&577 & 0&681 & 1&211\\ 
			\bottomrule
		\end{tabular}
\end{table} 
 
\begin{table}[htp]
	\caption{\label{table2.sim} 
		Simulation study results: posterior medians of total squared errors for the three functionals of interest and the three kernels.} 
	\centering
	\begin{tabular}{l r@{.}l r@{.}l r@{.}l r@{.}l r@{.}l r@{.}l r@{.}l r@{.}l r@{.}l}		
		& \multicolumn{6}{c}{$x < 3$} & \multicolumn{6}{c}{$3<x<6$} & \multicolumn{6}{c}{$6<x<9$}\\ 
		\midrule
		& \multicolumn{2}{c}{Mean}  & \multicolumn{2}{c}{Q$_{25}$} & \multicolumn{2}{c}{Q$_{75}$} 
		& \multicolumn{2}{c}{Mean}  & \multicolumn{2}{c}{Q$_{25}$} & \multicolumn{2}{c}{Q$_{75}$} 
		& \multicolumn{2}{c}{Mean}  & \multicolumn{2}{c}{Q$_{25}$} & \multicolumn{2}{c}{Q$_{75}$}\\		
		P       & 0&272  & 0&238 & 0&257 & 0&081 & 0&066 & 0&194 & 0&086 & 0&125 & 0&196\\  
		NB      & 0&143  & 0&114 & 0&161 & 0&056 & 0&053 & 0&119 & 0&038 & 0&304 & 0&297\\   
		GP      & 0&150  & 0&142 & 0&183 & 0&054 & 0&049 & 0&110 & 0&037 & 0&101 & 0&106\\  
		\cmidrule{1-19}
		& \multicolumn{6}{c}{$x>9$} & \multicolumn{12}{c}{} \\
		\cmidrule{1-7}
		& \multicolumn{2}{c}{Mean}  & \multicolumn{2}{c}{Q$_{25}$} & \multicolumn{2}{c}{Q$_{75}$}\\
		P & 0&267 & 0&295 & 0&869\\
		NB & 0&236 & 0&234 & 0&755\\
		GP & 0&289 & 0&314 & 0&763\\
		\bottomrule
	\end{tabular}
\end{table}

In Table \ref{table1.sim} we present the total errors. 
Concerning estimation of the mean surface, the model that utilises the NB distribution function performs the best, reducing the total errors
of the models that utilise the Poisson and GP distribution functions by $33\% (= 1-0.513/0.762)$ and $11\%$ respectively.
Concerning estimation of the quantiles, the best performance is achieved utilizing the GP function. Specifically,
estimation of the first quantile under the DPMM with the GP function is improved by $14\%$ and $20\%$ 
relative to the DPMMs that utilise the Poisson and NB distribution functions, while estimation of the third quantile is improved by $24\%$ and $15\%$,
respectively. 

In Table \ref{table2.sim} results are presented in more detail. 
Although the more detailed results are not always clear-cut, there are some general observations that can be made. 
Firstly, over the range $3<x<6$, where the response is over-dispersed relative to the Poisson, the DPMMs
with the NB and GP distribution functions perform better than the DPMM with the Poisson distribution function. 
The same is true also over the range $x<3$ where the response is Poisson distributed.
Secondly, estimation is substantially improved under the DPMM with the GP kernel over the region of the 
covariate space where the response is under-dispersed, $6<x<9$. Thirdly, over the region $x>9$, the DPMM 
with the NB distribution function does better that the DPMMs with the Poisson and GP distribution functions. 
It is a bit surprising that over $x>9,$ the DPMM with the Poisson distribution function does better than that with the GP distribution function 
for estimating the mean and first quantile functions, although the differences in estimation of the mean 
are mostly due to the results concerning estimation of conditional pmf $f(Y|X=x)$ for $x=10.9,$ which is at the edge of the covariate space.

Further, in Figure \ref{figure1.sim} we present a simulated dataset along with plots of the estimated mean and $25$th and $75$th percentile curves 
utilizing the GP distribution function. We can see that the model fits wells over all regions of the covariate space. 

\begin{figure}
	\begin{center}
		\begin{tabular}{ccc}
			\includegraphics[width=0.30\textwidth]{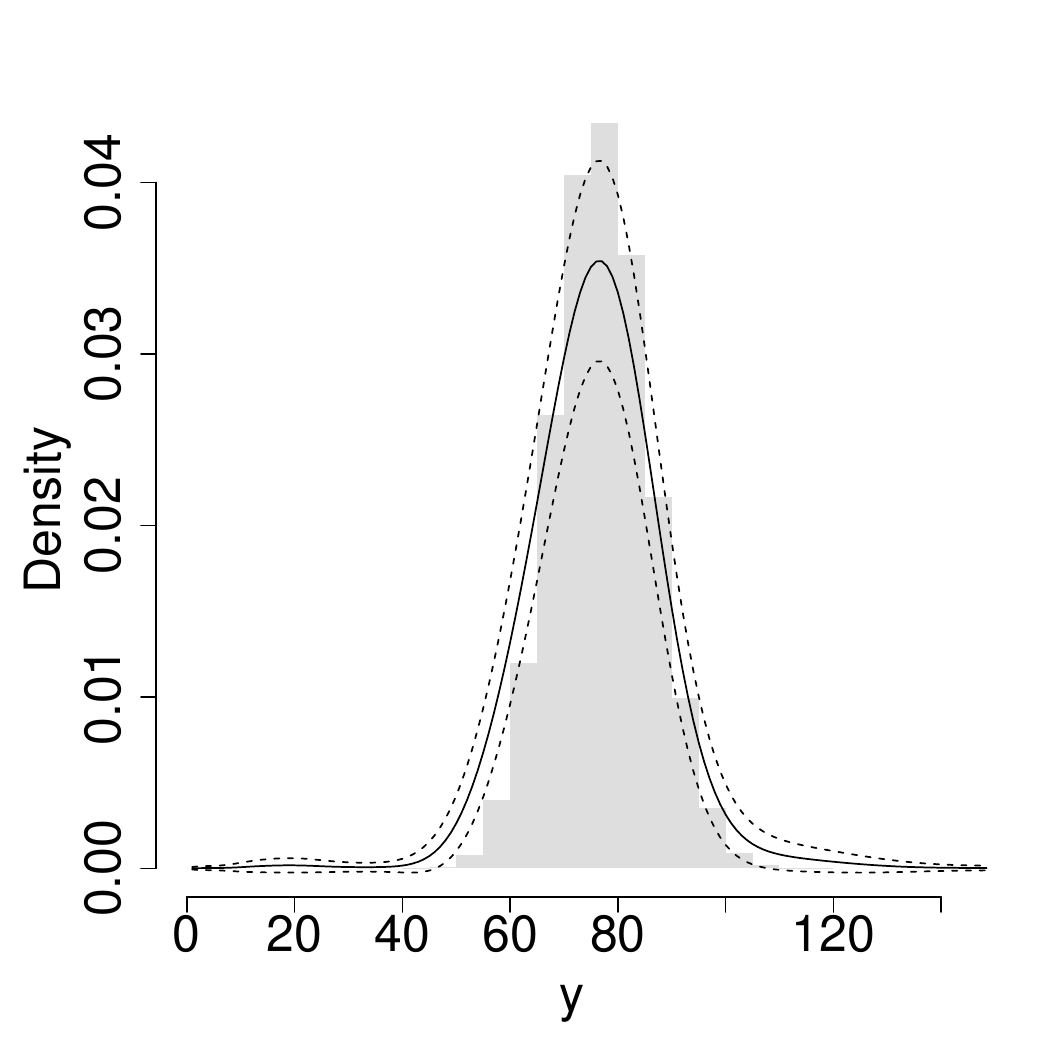} &
			\includegraphics[width=0.30\textwidth]{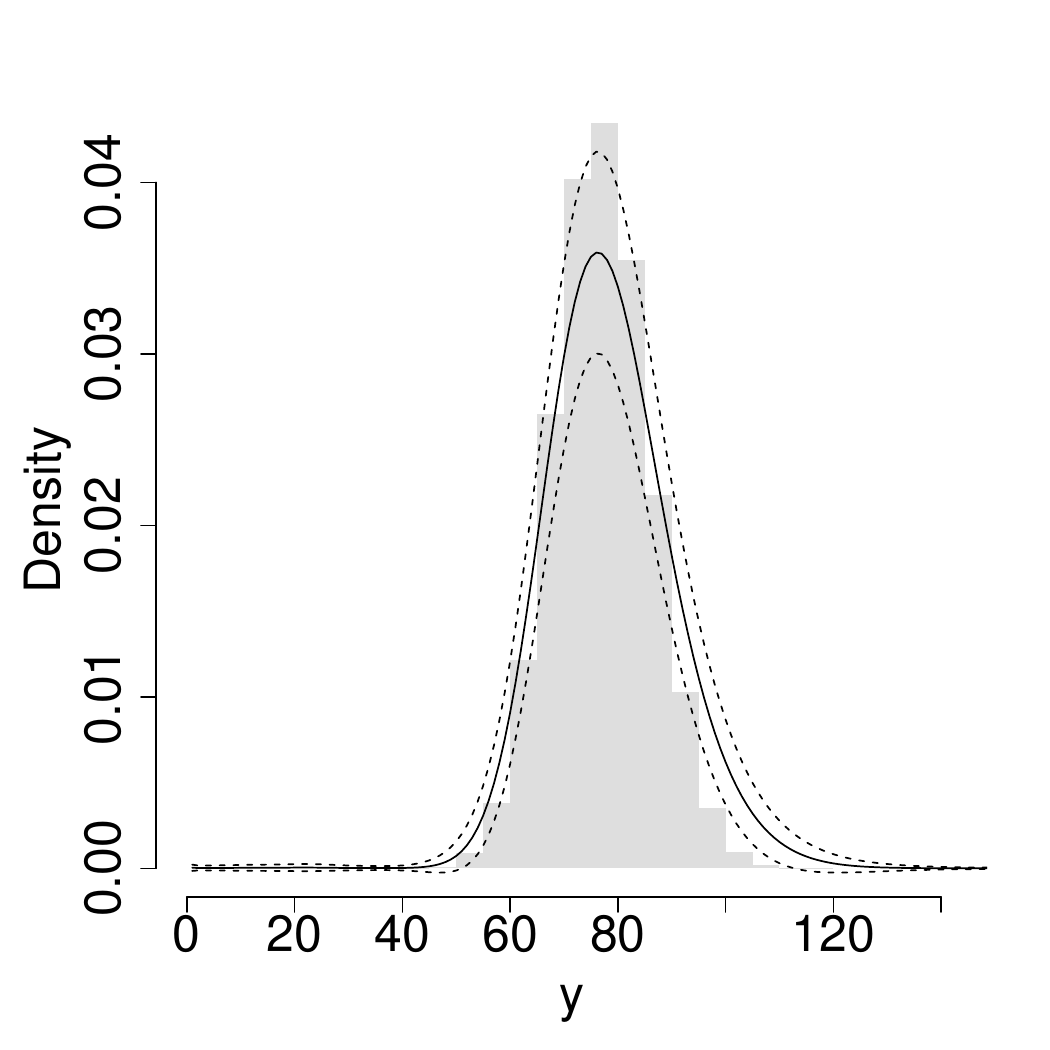} &
			\includegraphics[width=0.30\textwidth]{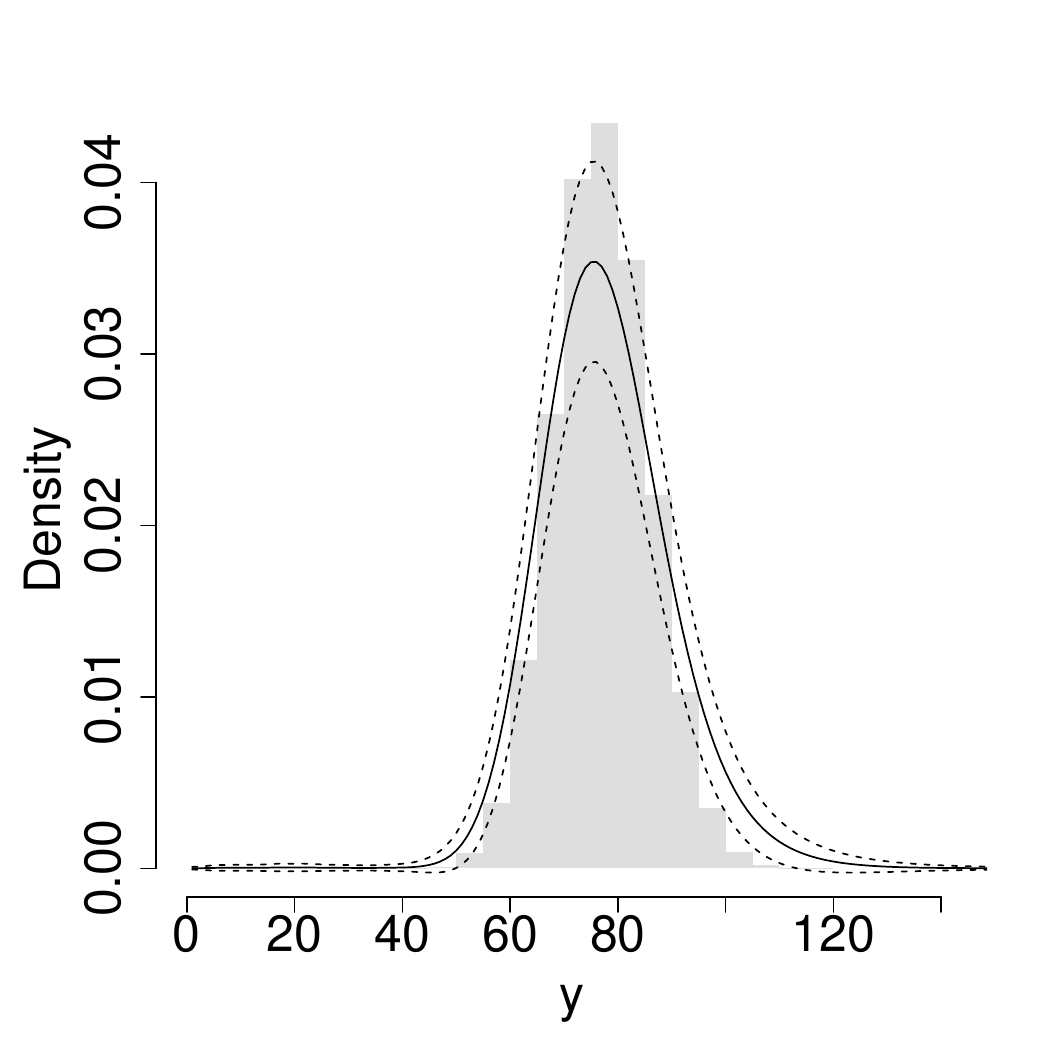}\\
			\includegraphics[width=0.30\textwidth]{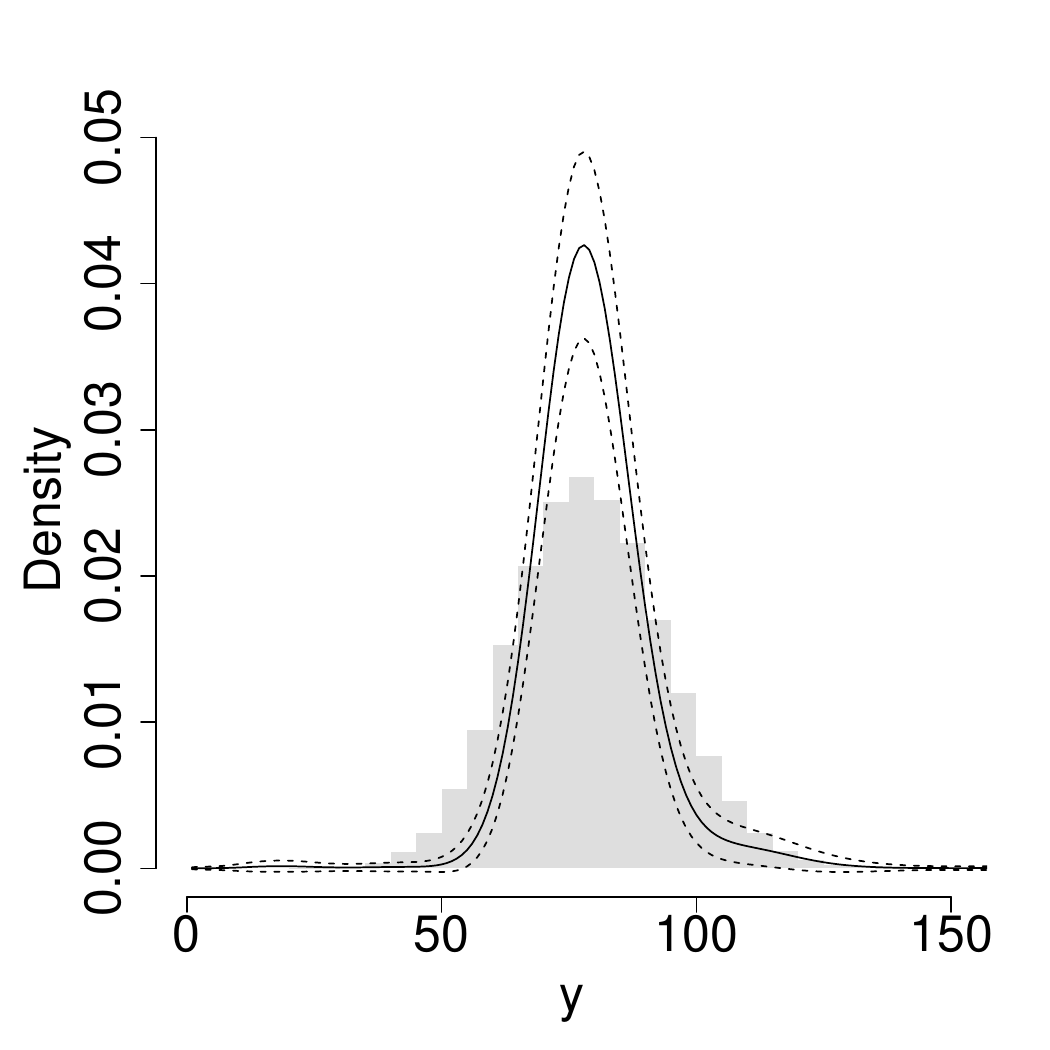} &
			\includegraphics[width=0.30\textwidth]{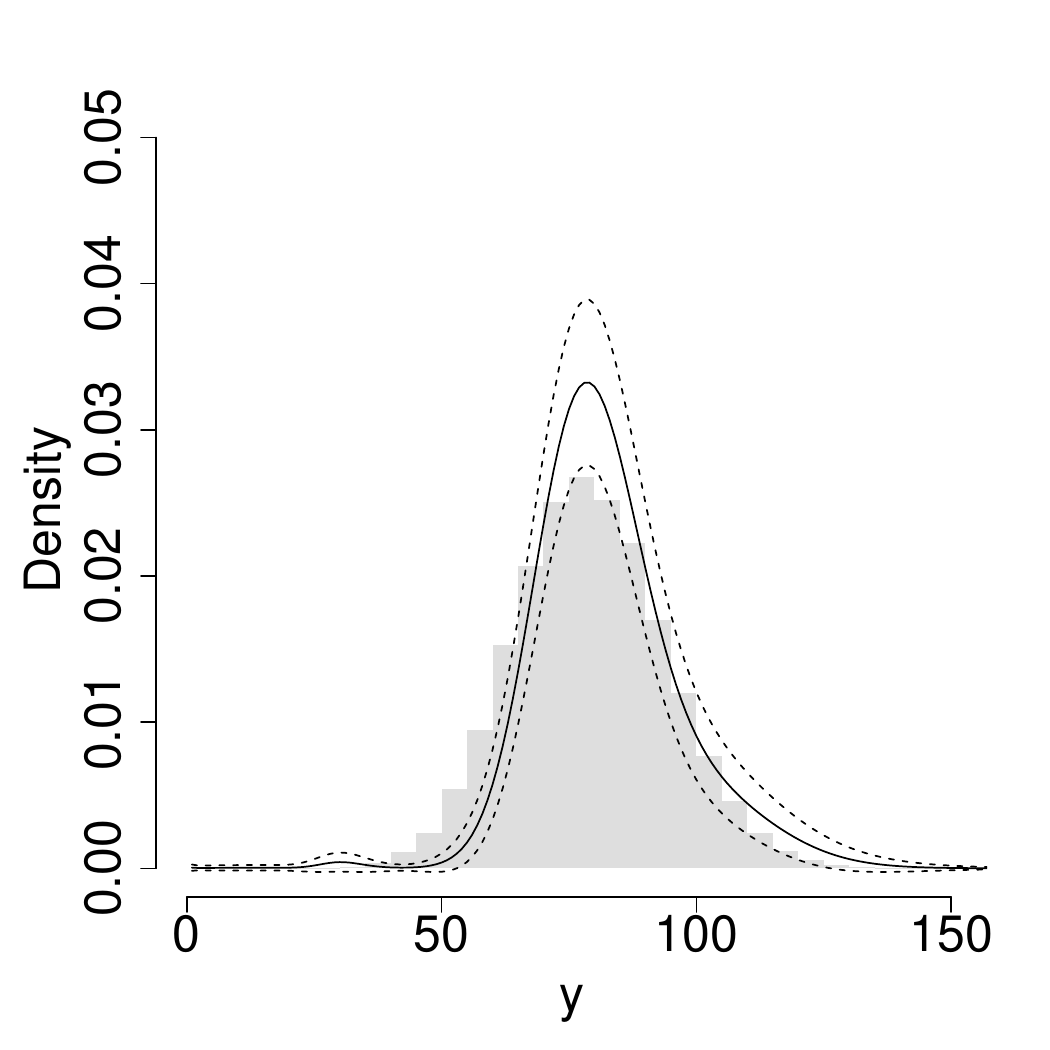} &
			\includegraphics[width=0.30\textwidth]{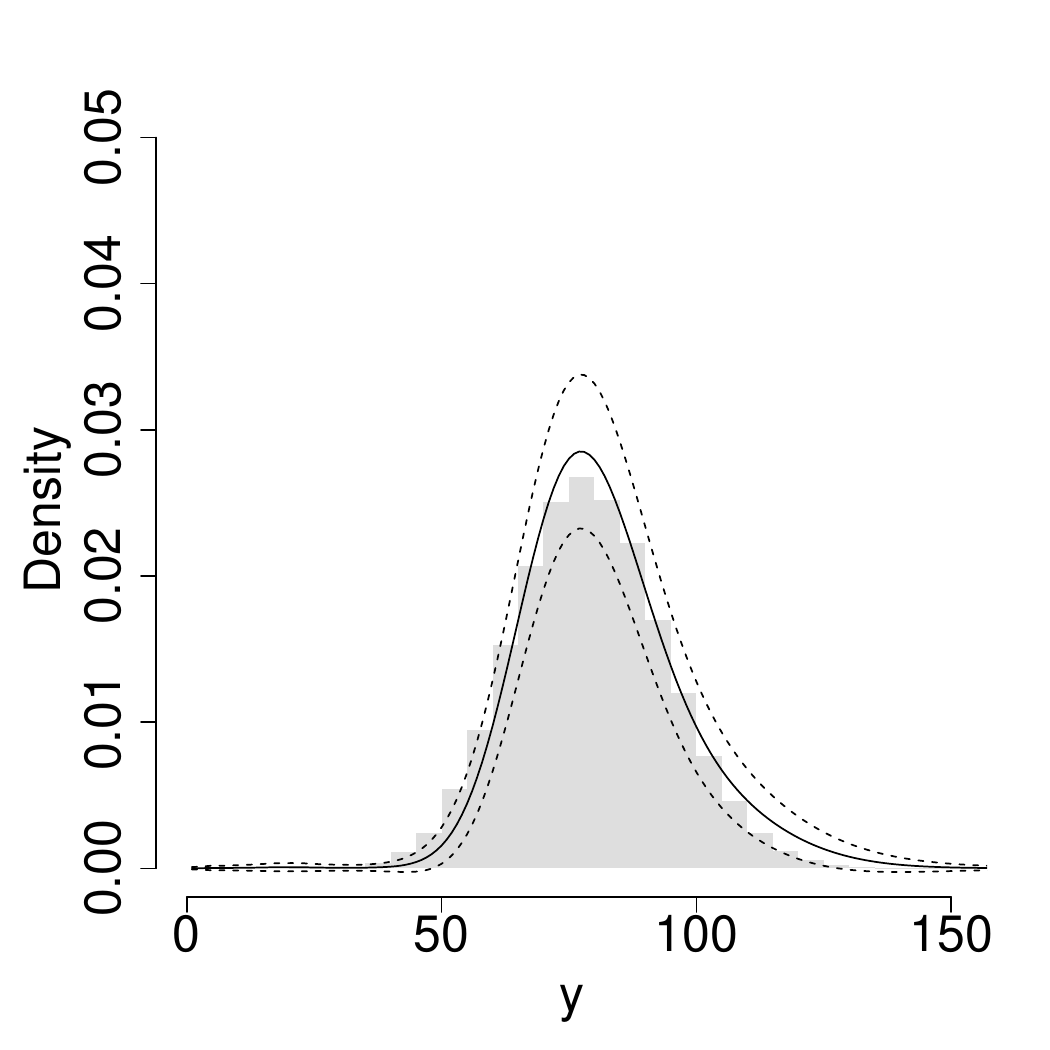}\\
			\includegraphics[width=0.30\textwidth]{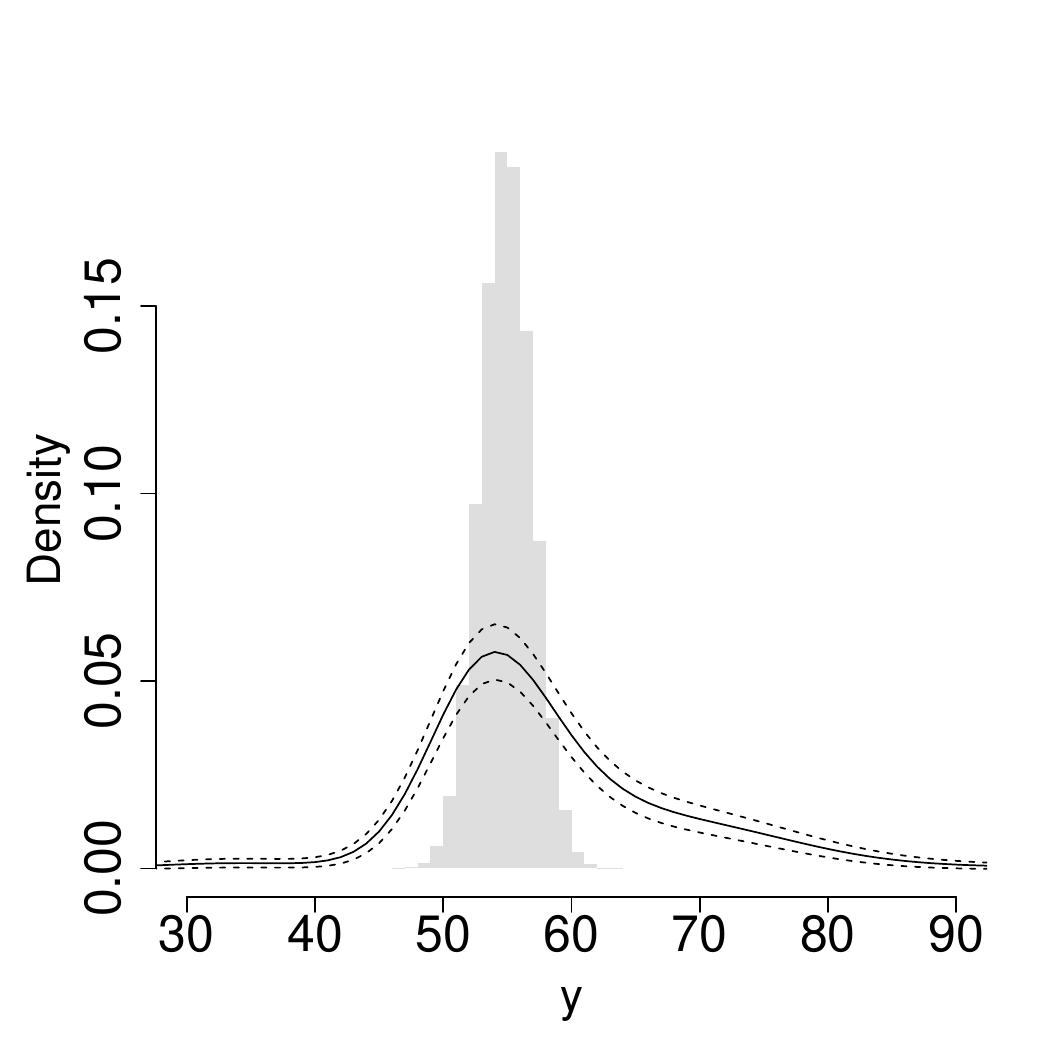} &
			\includegraphics[width=0.30\textwidth]{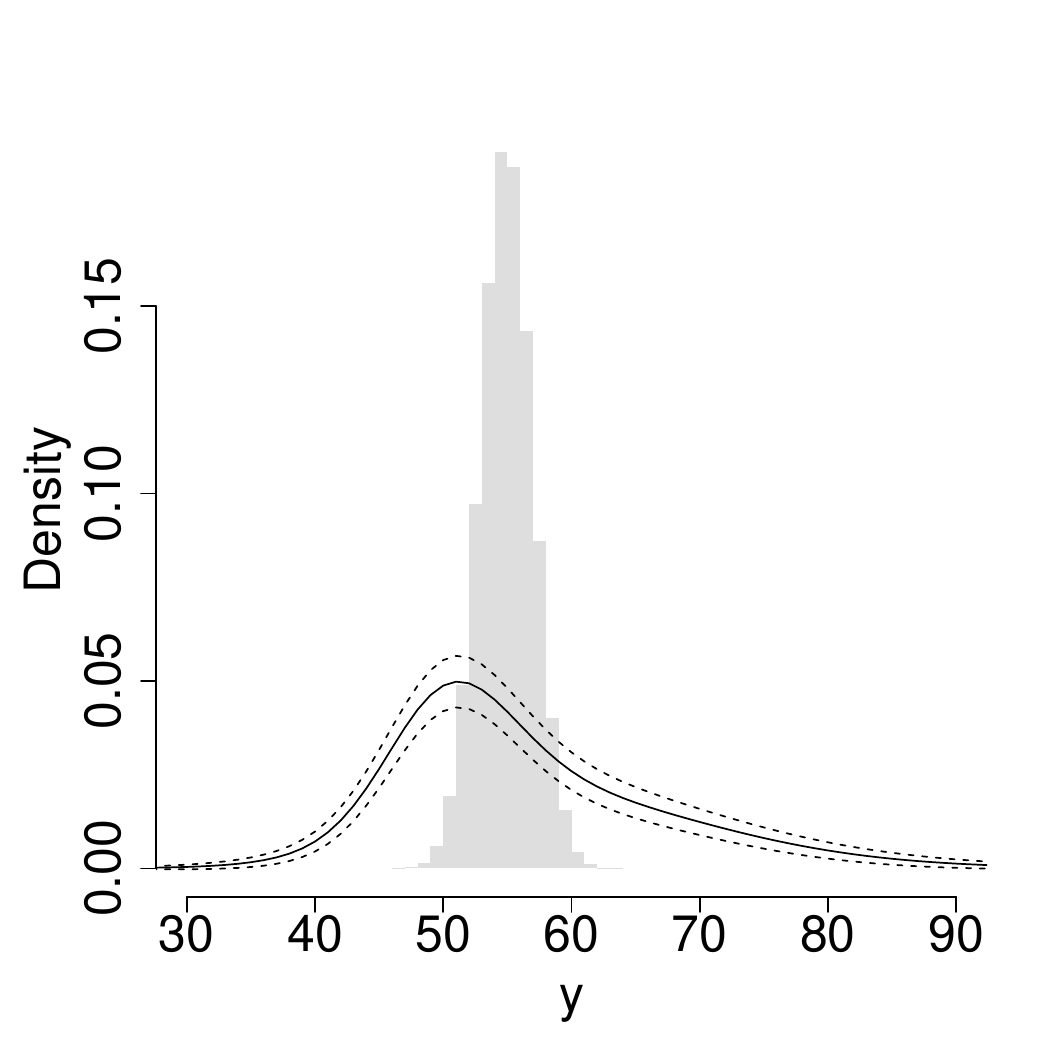} &
			\includegraphics[width=0.30\textwidth]{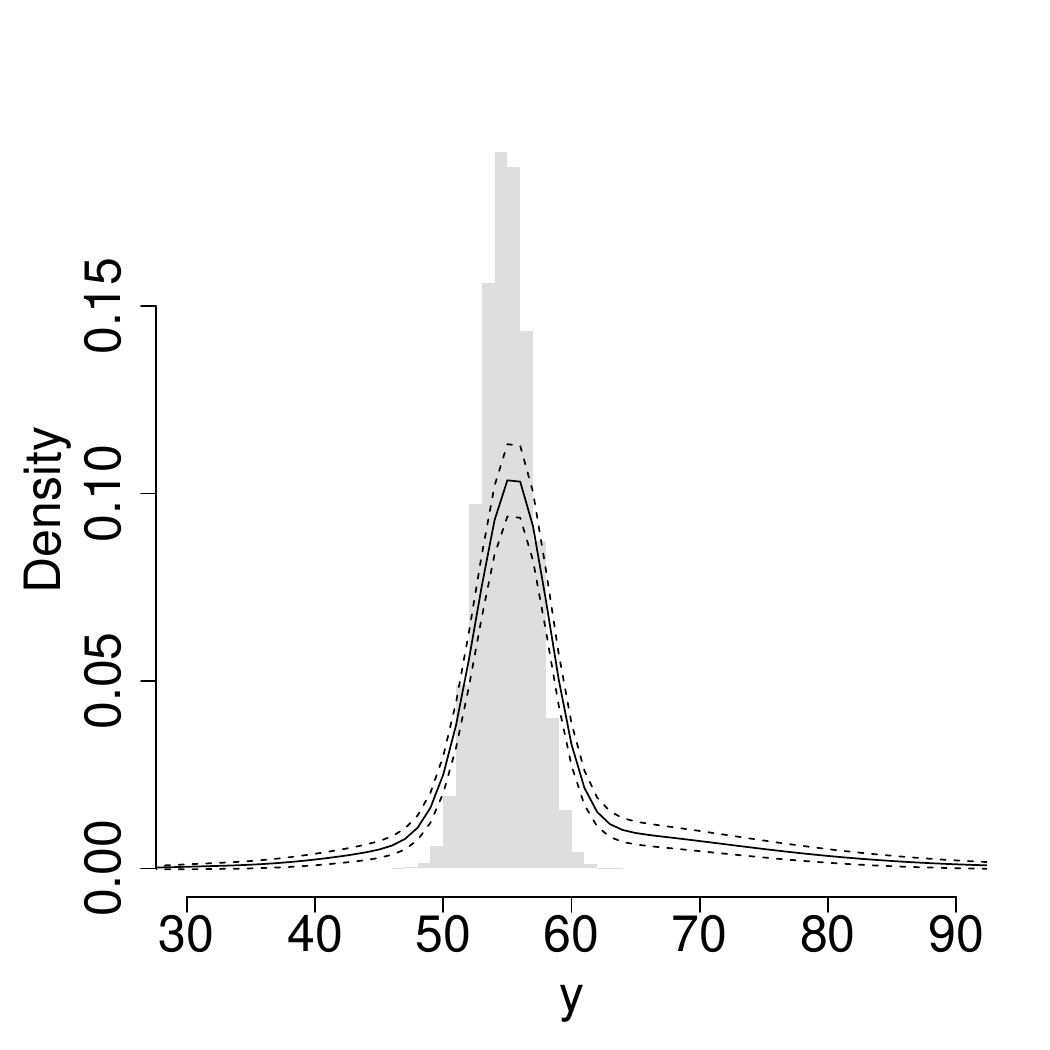}\\
		\end{tabular}
	\end{center}
	\caption{Estimated conditional pmfs $f(y|x)$ and 95\% credible intervals. The histograms represent data generated from the true model. 
		The rows correspond to $x=2.3, 3.7, 6.3$ where the response is equi-, 
		over- and under-dispersed relative to the Poisson. The three columns correspond to the models with Poisson, NB and GP distribution functions.}\label{figure2.sim}
\end{figure}

Figure \ref{figure2.sim} presents estimated conditional pmfs $f(y|x)$, and $95\%$ credible intervals, for three values of covariate $x$, namely
$x=2.3, 3.7, 6.3$, to show model performance over regions of the covariate space where the response is equi-, over- and under-dispersed relative
to the Poisson. The three rows of the figure correspond to the three values of $x$ and the three columns to the models with the 
Poisson, NB and GP distribution functions. In the first row, where the response is Poisson distributed, we see that all models fit well.
In the second row, where the response is over-dispersed, we see that the model that utilises the Poisson distribution function cannot adopt to the thicker tails. 
Lastly, in the third row, we see that only the model that utilises the GP function can adopt to the under-dispersion. 

\section{Application}\label{application}

\begin{figure}
	\begin{center}
		\includegraphics[width=0.45\textwidth]{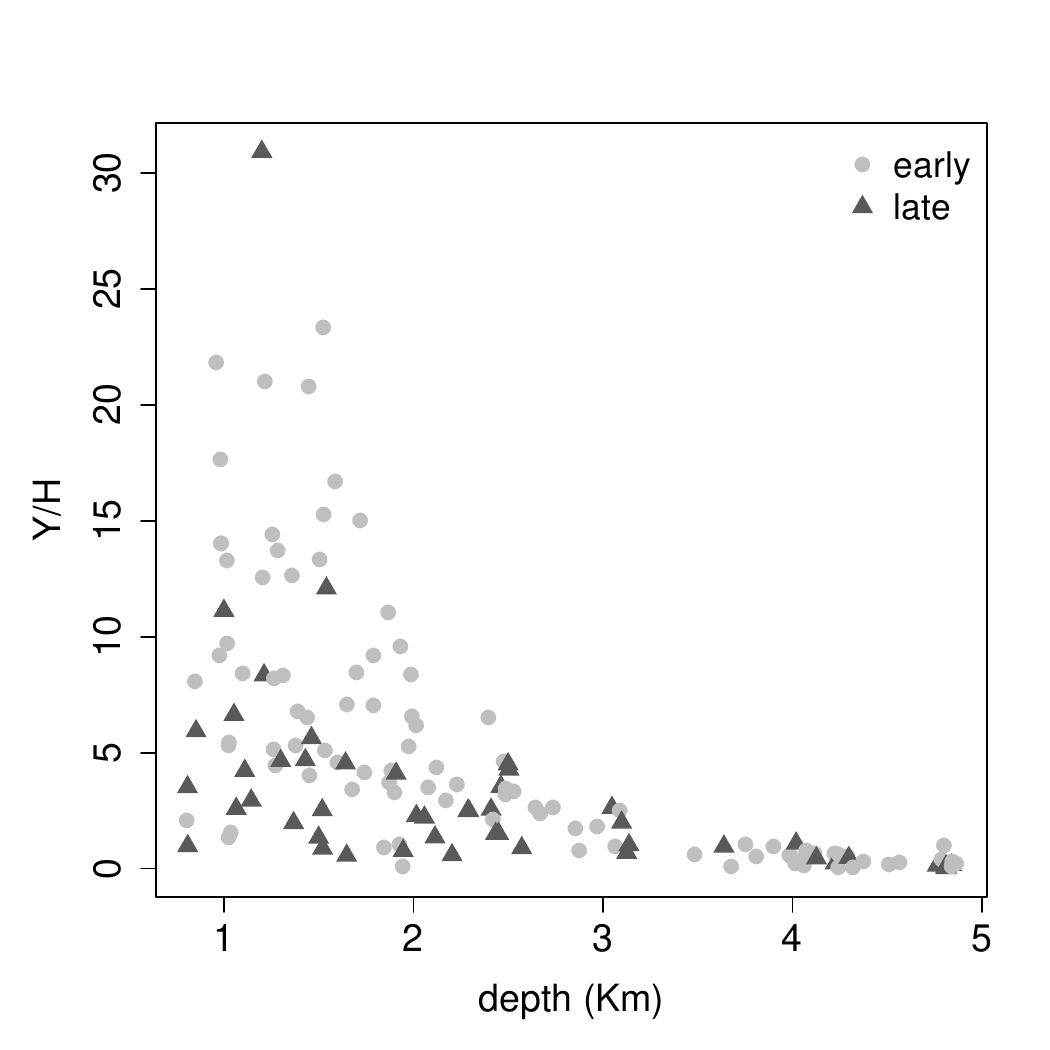}
	\end{center}
	\vspace{-.5cm} \caption{A scatter-plot of the fishing data. Standardised responses $(Y/H)$ plotted against
		depth ($x_2$) with different symbols for the two time periods ($x_1$).}
	\label{data1}
\end{figure}

We present an application that we adopt from \citet{Bailey09} who examined the 
impact of commercial fishing on deep-sea fish populations in the northeast Atlantic. 
The dataset, that is available in \citet{Hilbe}, includes $n=147$ observations on scientific 
trawls that were made in two 
distinct time periods, from $1977$ to $1989$ and from $1997$ to $2002$, at depths from $0.8$ 
to $4.8$ Km. The response variable $y$ is `fish abundance', a count of the number of fish 
caught in each of the $147$ trawls. With each trawl, there is an associated offset term $H$ 
that is calculated based on the size of the swept area (Km$^2$). The two explanatory variables 
are $x_1$, a time period indicator, and $x_2$, the average depth (Km) of the trawl. 
The main interest here is on the effect of the time periods that are thought to reflect the effect of the 
development of commercial fishing. The period $1977$--$1989$ (with $97$ observations),
which we refer to as the `early' period, is before and during the development of commercial fishing, 
and the period $1997$--$2002$ (with $50$ observations), which we refer to as the `late' period, 
is considered post-commercial fishing. The dataset is displayed in Figure \ref{data1}.

Commercial fishing is limited to depths of less than approximately $1.6$ Km. Hence, it may be
reasonable to expect fish abundance in waters deeper than $1.6$ Km to be unaffected by commercial fishing.
However, it is also possible that the effects of commercial fishing are transmitted to the deeper waters.
In fact, \citet{Bailey09} concluded that fish abundance reduced significantly between the two periods
at all depths between $0.8$ and $2.5$ Km. One of the two explanations they considered was that 
these reductions were due to commercial fishing and its effects cascading to deeper waters. Therefore,
we find it interesting to re-analyse the dataset using the methods we have described.

We let $\uz_i = (y_i,x_{i1},x_{i2})^{\top}, i=1,\ldots,n,$ denote the $i$th vector of observed response and covariates,
where the count of fish caught $y_i$ is associated with offset term $H_i$, the size of the swept area. 
The model we fit to $\uz_i, i=1,\ldots,n,$ is a DP mixture of the form
\begin{align*}
f_P(y_i,x_{i1},x_{i2}) = \sum_{h=1}^{T}\pi_h \int_{R(y_i)} \int_{R(x_{i1})} \text{N}(\uz_i^*;\umu_h^*,\uSigma_h^*)dy_i^* d\ux_{i1}^*,
\end{align*}
where $\uz_i^{*}=(y^*_i,x_{i1}^*,x_{i2})^{\top}$ is the vector of the latent and directly observed continuous variables.
Further, $\text{N}(\uz_i^*;\umu_h^*,\uSigma_h^*)$ denotes a trivariate Gaussian density with one restriction on the mean vector 
and two on the covariance matrix 
\begin{align}
\umu_h^* =
\begin{array}{ll}
\left[ 
\begin{array}{l}
0 \\
\mu_{1h} \\
\mu_{2h}
\end{array}
\right],\;
\uSigma_h^*=\left[ 
\begin{array}{lll}
1.0 &  \sigma_{12} & \sigma_{13}\\
\sigma_{21} &  1.0 & \sigma_{23}\\
\sigma_{31} &  \sigma_{32} & \sigma_{33}\\
\end{array}
\right]
\end{array}\label{exmpl2}.
\end{align}

The two rules for connecting observed and latent variables are as follows
\begin{align*}
&Y_i = y_i \text{\;\;if and only if\;\;} c_{y_i-1} < y_i^* \leq c_{y_i}, \text{where\;} c_{y_i} = \Phi^{-1}[F(y_i;H_i,\uxi)],\\
\nonumber\\
&X_{i1} = 0 \text{\;\;if and only if\;\;} x_{i1}^* < 0,
\end{align*}
where $F(;,)$, that appears in the definition of the cut-points, is taken to be the distribution function of a 
negative binomial random variable. This choice was guided by the presence of
over-dispersion and lack of under-dispersion in the response variable over the predictor space. 

To fit this model, we ran the MCMC algorithm for $60,000$ iterations, and retained one sample every five, after discarding the first $20,000$ as burn-in. Let $f_{P_s}(y_i,x_{i1},x_{i2})$ denote the $s$th sampled joint density, $s=1,\ldots,8,000$. From it, we can compute the $s$th sampled conditional pmf $f_{P_s}(y_i|x_{i1},x_{i2})$ that describes the possible values and associated probabilities for the count variable $y_i$, for the given values of the covariates, $x_{i1},x_{i2}$, and the given value of the offset term $H_i$, the size of the swept area. We sampled conditional pmfs for the early ($x_1=0$) and late ($x_1=1$) periods, in combination with $25$ equally spaced depths $(x_2)$, ranging from $0.90$ to $4.75$ Km, and with offset term $(H)$ 
fixed at the value of mean observed offset.     

This procedure enables inference about how the shape of the conditional pmf changes with covariates, 
which is what we refer to as pmf regression. Further,   
for each sampled conditional $f_{P_s}(y_i|x_{i1},x_{i2})$, we can calculate all functionals of interest. 
Here, we are interested in the median of the conditional pmf, and on how it changes with covariates,
which is what we refer to as median regression. Other functionals, such as the mode or quantiles of interest, can also 
be computed. This is an important feature of the approach presented here: whereas traditional methods,
such as generalised linear models, allow us to examine only how the mean changes with covariates, the current
method allows for more detailed examination of the effects of the covariates on the response distribution.  
  
\begin{figure}
	\begin{center}
		\begin{tabular}{cc}
			\includegraphics[width=0.45\textwidth]{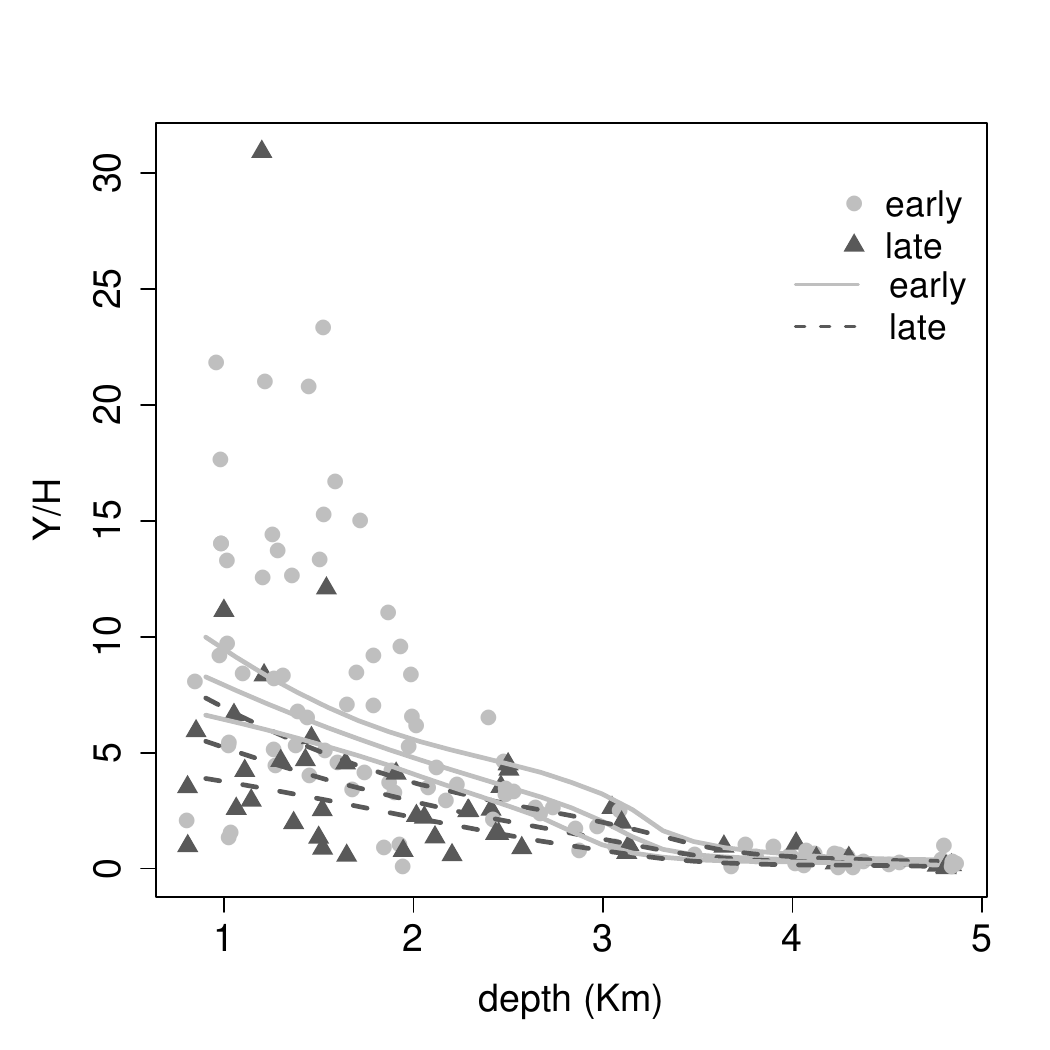} &
			\includegraphics[width=0.45\textwidth]{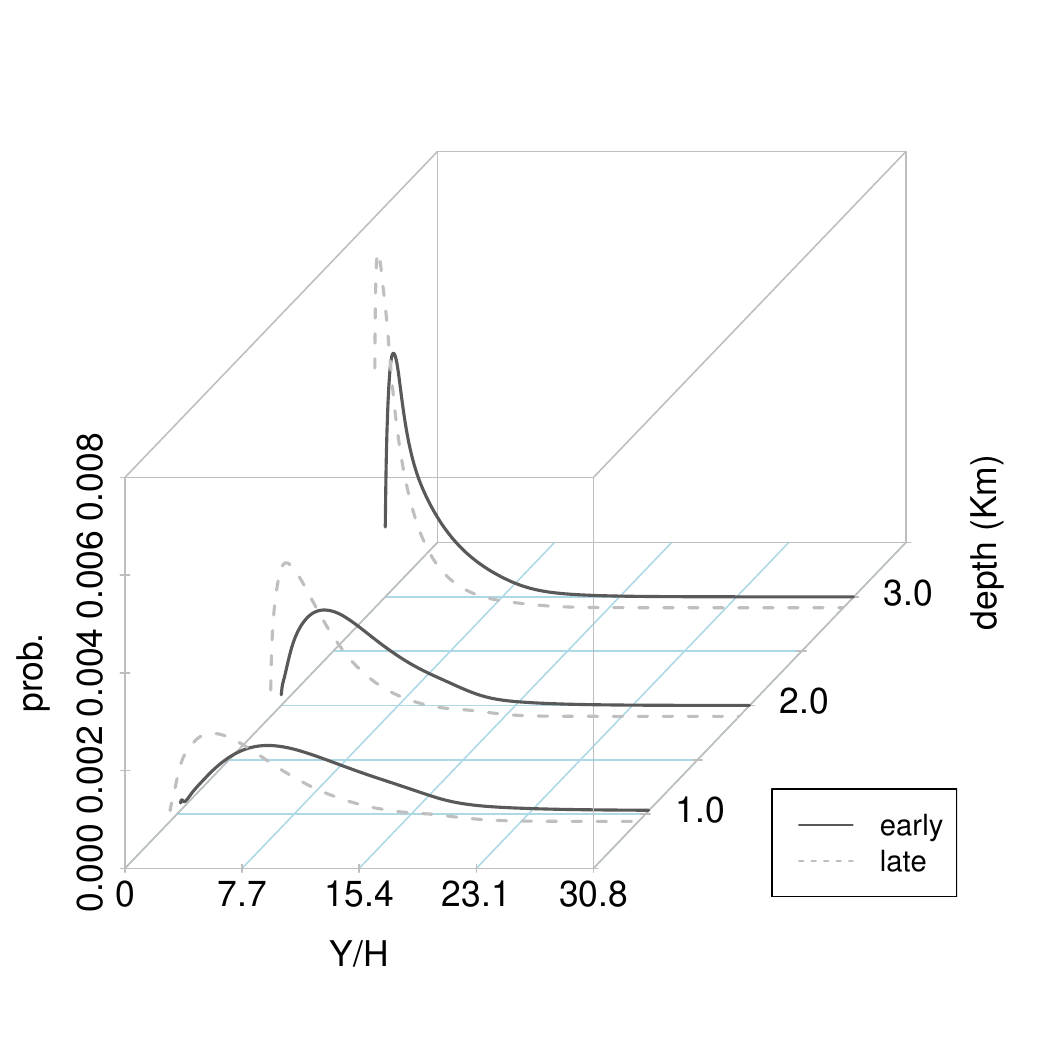} \\
			(a) & (b)\\
		\end{tabular}
	\end{center}
	\vspace{-.5cm} \caption{Results from the analysis of the fishing data. Figure (a) displays results on median
		regression along with $90\%$ credible intervals. Figure (b) displays results on pmf regression.}
	\label{results1}
\end{figure}

Results, in terms of standardised responses $Y/H$, are presented in Figure \ref{results1}. 
First, Figure \ref{results1} (a) displays the results for
median regression along with $90\%$ pointwise credible intervals. With solid lines are the results for the early 
period and with dashed lines the results for the late period. Clearly, median fish abundance decreases 
with depth for both periods. The median abundance for the late period is below that of the early period for all depths up to about $3.5$ Km. Credible intervals do not overlap for depths between about $1.3$ and $2.1$ Km. 
Second, Figure \ref{results1} (b) displays the results for pmf regression. With solid lines are the 
estimated pmfs for the early period and with dashed lines the ones for the late period. At depth $1.0$ Km, the 
pmf of the late period gives higher probability to smaller rates i.e. to smaller counts associated with offset 
term equal to unity. As the depth increases, the two pmfs give higher probability to smaller numbers, while
the estimated pmf of the late period continues to give higher probability to smaller numbers. 

\section{Discussion} \label{discussion}

We have developed Bayesian models for pmf regression with emphasis on count responses. 
The method represents discrete variables as continuous latent variables 
that have been discretised and utilises Dirichlet process mixtures of Gaussians
to model the joint density of the observed and latent continuous variables. 
The joint density forms the basis for carrying out 
inference on the conditional densities and its functionals.

The assumed mechanism by which latent continuous variables become observed discrete ones 
utilises cut-points that are expressed as $\Phi[F(;)]$, where $\Phi(\cdot)$ is the distribution function
of a standard normal variable and $F(;)$ is an appropriate distribution function, the choice of which
is made by the data analyst, depending on the needs of the particular data analysis that is being carried out. 
We have considered and evaluated several functions $F(;)$. 
We have shown utilizing simulated and real data how flexible the proposed model is and the diverse types of 
Bayesian inferences one can obtain by utilizing this model, including pmf, mean and quantile regression.
Another attractive feature of the current model is the ease by which missing data can be handled under a missing 
at random mechanism. See e.g. \citet{DB11} for further details. 

The method we have proposed can be computationally expensive. 
There are several parts of the MCMC algorithm that can contribute to that. 
First, obtaining samples of discrete distributions $f(y|\ux)$ over a set of covariate values $\ux$, 
to enable inference about the dependence of the pmf on the covariates,
is very computationally intensive, especially when a natural upper bound on the values of $y$ 
is not present. There are additional features in the model
that can make it computationally intensive. These are the restrictions that are placed on the model 
for the latent variables, the zero mean and unit variance, which are, of course, 
also present in the DP mixtures of Gaussians. 
Furthermore, the numerical integration over the unobserved latent variables can also be numerically 
intensive, depending on the number of discrete variables present in the model. 
Lastly, it is also computationally demanding to handle the possibly high dimensional covariance and 
precision matrices which are present in the model for the joint density. 
The MCMC algorithms for fitting the presented models, with any choice of the 
discussed functions $F(;)$, are available in the \textsf{R} package \textsf{BNSP} \citep{bnsp}.

\section{Appendix I} \label{mcmcA}

Our MCMC sampler proceeds as follows 
\begin{enumerate}
	\item Update $v_h \sim\text{Beta}(n_h+1,n-\sum_{l=1}^hn_h+\alpha),$
	where $n_h$ is the number of observations allocated in the $h$th cluster.
	
	\item The joint posterior of $(\uD_h, \uSigma_{h}^{*})$ is given by
	\begin{align*}
	p(\uD_h, \uSigma_{h}^{*}|\ldots) \propto |\uD_h|^{\eta/2-1} |\uSigma_{h}^*|^{(\eta-q-1-n_h)/2} 
	\exp[-\tr(\uEta^{-1} \uE_{h}+\uSigma_{h}^{*^{-1}} \uS_{h})/2],
	\end{align*}  
	where $\uS_h = \sum_{i:\delta_i=h}(\uz^*_i-\umu_h^*)(\uz^*_i-\umu_h^*)^{\top}$.
	
	As the above is a non-standard density, sampling from it requires a Metropolis-Hastings step.
	We take the proposal density to be 
	$\uE_h^{(p)}\sim$ Wishart$(\uE_h^{(p)};\psi,\uE_h^{(t)}/\psi)$, where 
	$\uE_h^{(t)}=\uD_h^{(t)^{1/2}} \uSigma_h^{*^{(t)}} \uD_h^{(t)^{1/2}}$
	and $\uD_h^{(t)}$, $\uSigma_{h}^{*^{(t)}}$ are realizations from the previous iteration. 
	Proposed values for $\uD_h, \uSigma_{h}^{*}$ are obtained by decomposing
	$\uE_h^{(p)}=\uD_h^{(p)^{1/2}} \uSigma_h^{*^{(p)}} \uD_h^{(p)^{1/2}}$
	and they are accepted with probability 
	\begin{align*}
	\alpha = \min\left\{\frac{p(\uD_h^{(p)}, \uSigma_{h}^{*^{(p)}}|\ldots)}{p(\uD_h^{(t)}, \uSigma_{h}^{*^{(t)}}|\ldots)}
	\frac{t(\uD_h^{(t)}, \uSigma_{h}^{*^{(t)}}|\uD_h^{(p)}, \uSigma_{h}^{*^{(p)}})}
	{t(\uD_h^{(p)},\uSigma_{h}^{*^{(p)}}|\uD_h^{(t)}, \uSigma_{h}^{*^{(t)}})},1\right\},
	\end{align*}
	where the proposal density is given by
	$t(\uD_h^{(p)},\uSigma_{h}^{*^{(p)}}|\uD_h^{(t)}, \uSigma_{h}^{*^{(t)}})
	= \text{Wishart}(\uE_h^{(p)};\psi,\uE_h^{(t)}/\psi)J(\uE_{h}^{(p)} \rightarrow \uD_h^{(p)}, \uSigma_{h}^{*^{(p)}})$.
	The free parameter $\psi$ is chosen adaptively \citep{roberts_examples_2009} so as to achieve an acceptance ratio of 
	about $20$--$25\%$ \citep{Roberts2001c}. Here the acceptance ratio that we adjust with parameter $\psi$ is the average 
	acceptance ratio over the non-empty clusters. 
	
	\item Next we describe how the means $\umu_{h}, h \geq 1$ are updated. 
	Recall that the $q$-dimensional mean $\umu_h^*$
	is restricted to have some of its elements equal to zero, see (\ref{restrictedMCV}). 
	These are the means that correspond to the latent variables underlying the discrete (but not the binary) variables. 
	Denote these by $\uz_1$ and their dimension by $p_1$. Further, the unrestricted elements of $\umu_h^*$, denoted by $\umu_h$, 
	correspond to the means of the latent variables underlying the binary and continuous variables. Denote these by $\uz_2$ and their 
	dimension by $p_2$, hence $p_1+p_2=q$. Writing the joint pdf of $(\uz_1^{\top},\uz_2^{\top})^{\top}$ as 
	\begin{align*}
	(\uz_1^{\top},\uz_2^{\top})^{\top} |(\umu_h^*,\uSigma^*_h)  \sim \text{N}\left(
	\umu_h^*=
	\begin{array}{cc}
	\left[ 
	\begin{array}{l}
	\uzero\\
	\umu_{h} \\
	\end{array}
	\right],
	&
	\uSigma^*_h=\left[ 
	\begin{array}{ll}
	\uSigma_{11h} & \uSigma_{12h}\\
	\uSigma_{21h} & \uSigma_{22h}\\
	\end{array}
	\right]
	\end{array}\right),
	\end{align*}
	it is easy to see that $\umu_h$ are updated from $f(\umu_{h}|\ldots) \propto$ 
	\begin{align}	 
	p_0(\umu_{h}) \prod_{\{i: l_{i}=h\}} \text{N}(\uz_{2i};\umu_{h}+\uSigma_{21h} \uSigma_{11h}^{-1} \uz_{1i}, 
	\uSigma_{22h} - \uSigma_{21h} \uSigma_{11h}^{-1} \uSigma_{12h}). \label{cond1}
	\end{align}
	With prior $\umu_{h} \sim \text{N}(\ud,\uD)$, the updating distribution is
	\begin{align*}
	\text{N}\Big\{&
	(n_h\uW^{-1}_h+\uD^{-1})^{-1}\big[\uW_h^{-1} \sum_{l_{i}=h} \left(\uz_{2i}-\uSigma_{21h} \uSigma_{11h}^{-1} \uz_{1i}\right) + \uD^{-1} \ud\big],\nonumber\\ 
	&(n_h\uW^{-1}_h+\uD^{-1})^{-1}\nonumber
	\Big\},
	\end{align*}
	where $\uW_h = \uSigma_{22h} - \uSigma_{21h} \uSigma_{11h}^{-1} \uSigma_{12h}$.
	
	\item We describe the step for updating $\uxi_{h}$, $h \geq 1,$ assuming that the dataset consists of binary and continuous covariates
	and a count or a binomial response variable. We update $\uxi_{h}$, $h \geq 1,$ from the marginal posterior having 
	integrated out $y^*$, the latent variable that corresponds to the response
	\begin{align*}
	p(\uxi_{h}|\ldots) \propto \prod_{\{i:\delta_{i}=h\}}\Big(\Phi\big\{[c_{y_{i}}(\ulambda_{i})-\text{E}_{i}^*]/\text{sd}_{i}^*\big\} -
	\Phi\big\{[c_{y_{i}-1}(\ulambda_{i})-\text{E}_{i}^*]/\text{sd}_{i}^*\big\}\Big) p_0(\uxi_{h}),
	\end{align*}
	where $\text{E}_{i}^* = \text{E}(y_{i}^{*}|\ux_{d,i}^*,\ux_{c,i})$ is the conditional expectation and
	$\text{sd}_{i}^* = \text{sd}(y_{i}^*|\ux_{d,i}^*,\ux_{c,i})$ is the conditional standard deviation. They are obtained using standard theory on multivariate normal densities, as was done in (\ref{cond1}). 
	Further, priors $p_0(\uxi_{h})$ are defined in Table \ref{fam2}.
	
	For all cases, updating requires a Metropolis-Hastings step. We provide details next.  
	\begin{enumerate}
		\item \label{vehicle} For Poisson mixtures, the proposed value $\xi_h^{(p)}$ is obtained from 
		$\xi_h^{(p)} \sim \text{Gamma}(\tau \xi_h^{(t)}\xi_h^{(t)},\tau \xi_h^{(t)})$, that is from a gamma 
		distribution with mean $\xi_h^{(t)}$ and variance $1/\tau$, where $\xi_h^{(t)}$ denotes the current value.   
		The acceptance probability is given by 
		\begin{align} 
		\min\{1, \frac{p(\xi_{h}^{(p)}|\ldots) \text{Gamma}(\xi_{h}^{(t)};\tau \xi_{h}^{(p)}\xi_{h}^{(p)},\tau \xi_{h}^{(p)})}{ 
			p(\xi_{h}^{(t)}|\ldots) \text{Gamma}(\xi_{h}^{(p)};\tau \xi_{h}^{(t)}\xi_{h}^{(t)},\tau \xi_{h}^{(t)})}\}. \label{ap}
		\end{align}
		Here $\tau$ is introduced as a free parameter which is adjusted adaptively \citep{roberts_examples_2009} in order to achieve 
		an acceptance ratio of about $20$--$25\%$ \citep{Roberts2001c}. The acceptance ratio that we adjust with parameter $\tau$ is the average 
		acceptance ratio over the non-empty clusters.
		
		\item For binomial mixtures, the proposed value $\xi_h^{(p)}$ is obtained from 
		$\xi_h^{(p)} \sim \text{Beta}(\xi_h^{(p)};a^{(t)},b^{(t)})$, where 
		$b^{(t)} = \xi_h^{(t)}-1+\xi_h^{(t)}(1-\xi_h^{(t)})^2\tau$ 
		and
		$a^{(t)} = b^{(t)} \xi_h^{(t)}/(1-\xi_h^{(t)})$ 
		that define a beta distribution with mean $\xi_h^{(t)}$ and variance $1/\tau$.    
		The expression of the acceptance probability follows along the same lines as (\ref{ap}) and hence omitted.
		
		\item\label{h3} For negative binomial mixtures, parameter vector $\uxi_h = (\xi_{1h},\xi_{2h})^{\top}$ is updated in a single step. 
		Proposed values for the elements of $\uxi_{h}^{(p)} = (\xi_{1h}^{(p)},\xi_{2h}^{(p)})^{\top}$ are obtained from 
		independent gamma distributions similar in form to the gamma distribution shown in part \ref{vehicle}
		for Poisson mixtures. We utilise a common tuning parameter $\tau$ in the two gamma proposal distributions.
		
		\item For beta-binomial mixtures, the elements of $\uxi_h = (\xi_{1h},\xi_{2h})^{\top}$ are also updated in a single step. 
		Proposed values for $\uxi_{h}^{(p)} = (\xi_{1h}^{(p)},\xi_{2h}^{(p)})^{\top}$ are obtained from 
		independent gamma distributions with a single tuning parameter, as was done in part \ref{h3}.  
		
		\item For generalised Poisson mixtures, parameter vector $\uxi_h = (\xi_{1h},\xi_{2h})^{\top}$ is updated in two steps
		utilizing two tuning parameters, $\tau_1$ and $\tau_2$. 
		Proposed values for the mean $\xi_{1h}$ are obtained from $\xi_{1h}^{(p)} \sim \text{Gamma}(\tau_1 \xi_h^{(t)}\xi_h^{(t)},\tau_1 \xi_h^{(t)})$
		and those for the variance $\xi_{2h}$ from $\xi_{2h}^{(p)} \sim \text{N}(\xi_{2h}^{(t)},1/\tau_2)$, a normal distribution centered at the 
		previous realization and with variance $1/\tau_2$. 
		
	\end{enumerate}
	
	\item We impute the latent variables $y_{i,}^*, \ux^*_{d,i}=(x^*_{d,i,1},\ldots,x^*_{d,i,p_d})^{\top}, i=1,\ldots,n,$ from 
	the conditional
	\begin{align*} 
	& (y_{i,}^*, \ux^{*^{\top}}_{d,i})^{\top}|\ux_{c,i} \sim
	\text{N}(y_i^*,\ux_{d,i}^*|\ux_{c,i};\umu_{h,d.c,i},\uSigma_{h,d.c})  
	\uone[y_{i}^{*} \in R(y_i)] \nonumber\\
	& \times \prod_{m=1}^{p_d} \uone[x_{d,i,m}^{*} \in R(x_{d,i,m})],
	\end{align*}
	where $\umu_{h,d.c,i}$ and $\uSigma_{h,d.c}$ were defined after (\ref{num1}). 
	The imputation utilises the algorithm of \citet{Robert2009} according to which 
	imputation is done one variable at a time given all other ones.
	Here with subscript $h$ we denote the cluster in which sampling unit $i$ is allocated. 
	
	\item We update the cluster allocation variables $\delta_{i},i=1,2,\ldots,n,$ according to probabilities $\Pr(\delta_{i}=h)$ obtained from the marginalised posterior
	\begin{align*}
	k_{ih} \text{N}(\ux_{c,i};\umu_{2h},\uSigma^*_{22h})
	\int \int \ldots \int \text{N}(y_i^*,\ux_{d,i}^*|\ux_{c,i};\umu_{h,d.c,i},\uSigma_{h,d.c})dy_{i,}^* d\ux^*_{d,i},
	\end{align*}
	where 
	$k_{ih}=\uone(u_i < \pi_{h})$ for the slice sampler while $k_{ih}=\pi_{h}$ for the truncated sampler.
	
	\item Label switching moves \citep{PR08}: 
	\begin{enumerate}
		\item[(a)] Choose randomly two nonempty clusters, $a$ and $b$ say, and propose to exchange their labels.  
		The acceptance probability of this move is
		$\min \left\{1, (\pi_{b}/\pi_{a})^{n_a-n_b} \right\}$.
		If the proposed move is accepted, we exchange allocation variables and cluster specific parameters.
		
		\item[(b)] Choose randomly a cluster, $a$ say, and propose to exchange the labels of clusters 
		$a$ and $a+1$, and at the same time propose to exchange $v_a$ with $v_{a+1}$. 
		Cluster $a$ is chosen randomly among clusters labelled
		$1,\ldots,n^*-1$, where $n^*$ is the nonempty cluster with the largest label.
		The acceptance probability of this move is
		$\min \left\{1,(1-v_{a+1})^{n_a}/(1-v_{a})^{n_{a+1}}\right\}$, and if 
		it is accepted, we exchange allocation variables and cluster specific parameters. 
	\end{enumerate}
	
	\item We update concentration parameter $\alpha$ using the method described by \citet{EscobarWest}.
	Assuming a Gamma$(\alpha|a,b)$ prior (mean $=a/b$), the posterior is expresses as a 
	mixture of two gamma distributions: 
	\begin{align}\label{alphadist}
	\alpha \sim \pi_{\eta} \text{Gamma}(a+k,b-\log(\eta)) + (1-\pi_{\eta}) 
	\text{Gamma}(a+k-1,b-\log(\eta)),
	\end{align}
	where $k$ is the number of non-empty clusters, 
	$\pi_{\eta} = (a+k-1)/\{a+k-1+n[b-\log(\eta)]\}$ and 
	\begin{align}\label{etadist}
	\eta|\alpha,k \sim \text{Beta}(\alpha+1,n). 
	\end{align}
	Hence the algorithm proceeds as follows: with $\alpha$ and $k$ fixed at their current values, we sample $\eta$ from
	(\ref{etadist}). Then, based on the same $k$ and the value of $\eta$, we sample a new $\alpha$ value
	from (\ref{alphadist}). 
	
\end{enumerate}

\section{Appendix II} \label{wcp}

We start by constructing a density $f_{p_{\epsilon}}$ such that 
$\sum_y \int_{x} f_0(y,\ux)\log[f_0(y,\ux)/f_{p_{\epsilon}}(y,\ux)]d\ux < \epsilon$ for any $\epsilon > 0$.

Let
\begin{align*}
dP_m(\uxi,\umu,\unu,\uSigma) = \delta_{\nu}(\uzero) \delta_{\Sigma}(h_m^2 \uI) f_0^*(\uxi,\umu),
\end{align*}
where $h_m = m^{-\eta}$ for some $\eta>0$ and $\delta_x(A) = \uone(x \in A)$.

Hence, by utilizing the kernel in (\ref{kernel2}), we may write 
\begin{align*}
f_{p_m}(y,\ux) = \int k(\uz;\utheta) dP_m(\utheta) = \int \int \text{N}(\ux;\umu,h^2_m \uI) K(y;\uxi) f_0^*(\uxi,\umu) d\umu d\uxi,
\end{align*}
where $K(y;\uxi)$ is the chosen model e.g. the Poisson, negative binomial or generalised Poisson pmf for count data. 

Now, by utilizing the transformation $\ua = (\ux-\umu)/h_m$, we obtain
\begin{align*}
f_{p_m}(y,\ux) = \int \int \text{N}(\ua;\uzero,\uI) K(y;\uxi) f_0^*(\uxi,\ux-h_m\ua) d\ua d\uxi.
\end{align*}
By the continuity of $f_0^*$, we have that $f_0^*(\uxi,\ux-h_m\ua) \rightarrow f_0^*(\uxi,\ux)$ 
as $h_m \rightarrow 0$. Further, recalling that by condition $C_2$, $f_0^*$
is bounded, by the dominated convergence theorem we have that 
\begin{align*}
f_{p_m}(y,\ux) \rightarrow \int K(y;\uxi) f_0^*(\uxi,\ux) d\uxi = f_0(y,\ux),
\end{align*}
where the last equality follows from condition $C_1$.
Therefore, as $m \rightarrow \infty$, $\log[f_0(y,\ux)/f_{p_m}(y,\ux)] \rightarrow 0$ for all $y$ and $\ux$. To show that 
\begin{align}
\sum_{y} \int_{\ux} f_0(y,\ux) \log[f_0(y,\ux)/f_{p_m}(y,\ux)] d\ux \rightarrow 0, \label{goal}
\end{align}
we need to find a function that dominates $|\log[f_0(y,\ux)/f_{p_m}(y,\ux)]|$ and that is $f_0$-integrable.

To this end, first observe that due to condition $C_2$, $f_{p_m}(y,\ux)$ is bounded from above by 
\begin{align*}
f_{p_m}(y,\ux) \leq M \int \int \text{N}(\ux;\umu,h^2_m \uI) K(y;\uxi) d\umu d\uxi \leq M.
\end{align*}
It follows that 
\begin{align}
\log \frac{f_0(y,\ux)}{f_{p_m}(y,\ux)} \geq \log \frac{f_0(y,\ux)}{M}. \label{comb1}
\end{align}

Further, for $||\ux||>m$ and any $y$, from $C_1$ and $C_4$, we have
\begin{align}
f_{p_m}(y,\ux) &\geq \int \int_{||\umu||<||\ux||} \text{N}(\ux;\umu,h_m^2\uI) K(y;\uxi) f_0^*(\uxi,\umu) d\umu d\uxi \nonumber\\
&\geq f_0(y,\ux) \int_{||\umu||<||\ux||} \text{N}(\ux;\umu,h_m^2\uI) d\umu \geq f_0(y,\ux)/3, \label{lim1}
\end{align}
where the last inequality follows by a suitable choice of $h_m^2$.   

Furthermore, for $||\ux||\leq m$ and any $y$, from $C_1$ and $C_3$, we have 
\begin{align}
f_{p_m}(y,\ux) &\geq  \int \int_{||\umu||<m} \text{N}(\ux;\umu,h_m^2\uI) K(y;\uxi) f_0^*(\umu|\uxi) f_0^*(\uxi) d\umu d\uxi \nonumber\\
&\geq c f_0(y) \int_{||\umu||<m} \text{N}(\ux;\umu,h_m^2\uI) d\umu \geq c f_0(y)/3. \label{lim2}
\end{align}

Combining (\ref{lim1}) and (\ref{lim2}) 
\begin{align*}
f_{p_m}(y,\ux) \geq \Big\{
\begin{array}{ll}
f_0(y,\ux)/3, & ||\ux||>m, \\
c f_0(y)/3, & ||\ux||\leq m, \\
\end{array}
\end{align*}
from which follows that 
\begin{align}
\log \frac{f_0(y,\ux)}{f_{p_m}(y,\ux)} \leq \xi(y,\ux) = \Big\{
\begin{array}{ll}
\log 3, & ||\ux||>m, \\
\log\{[3 f_0(y,\ux)]/[c f_0(y)]\}, & ||\ux|| \leq m.  \label{comb2}
\end{array}
\end{align}

Now, from (\ref{comb1}) and (\ref{comb2})
\begin{align*}
\left|\log \frac{f_0(y,\ux)}{f_{p_m}(y,\ux)}\right| \leq 
\max\left\{\xi(y,\ux),\left|\log \frac{f_0(y,\ux)}{M}\right|\right\},
\end{align*}
where the right-hand side is $f_0$-integrable due to $C_5$, and it follows that (\ref{goal}) holds.

For any given $\epsilon > 0$, condition $A_1$ is satisfied by $f_{P_m}$ with suitable choice of $m$. 
Hence, we take $f_{P_{\epsilon}} = f_{P_m}$. 

Further, to show that condition $A_2$ is satisfied, observe that 
\begin{align}
c = \inf_{\uz \in \mathcal{Z}} \inf_{\utheta \in \uTheta} k(\uz;\utheta) > 0, \label{elev}
\end{align}
where $\mathcal{Z}$ denotes the sample space. 

In addition, note that the family of maps $\{ \utheta \rightarrow k(\uz;\utheta): \uz \in \mathcal{Z} \}$ is uniformly 
equicontinuous over compact space $\uTheta$.
To see this, write
\begin{align*}
& k(\uz;\utheta) = \int_{R(y)} \text{N}(\uz^*;\umu^*,\uSigma^*) dy^{*} = \\
& \text{N}(\ux;\umu,\uSigma) \int_{R(y)}  \text{N}(y^{*};m^*,v^*) dy^{*} = 
\text{N}(\ux;\umu,\uSigma) p(y;\ux,\utheta),
\end{align*} 
where $m^*$ and $v^*$ were defined below (\ref{kernel2}).
Now, $|k(\uz;\utheta) - k(\uz;\utheta^{\prime})|$ is expressed as
\begin{align}
&|\text{N}(\ux;\umu,\uSigma) p(y;\ux,\uxi,\umu,\uSigma)-\text{N}(\ux;\umu^{\prime},\uSigma^{\prime}) p(y;\ux,\uxi^{\prime},\umu^{\prime},\uSigma^{\prime})| \\
& \leq p(y;\ux,\uxi,\umu,\uSigma) |\text{N}(\ux;\umu,\uSigma)-\text{N}(\ux;\umu^{\prime},\uSigma^{\prime})| + \nonumber\\
& \text{N}(\ux;\umu^{\prime},\uSigma^{\prime})|p(y;\ux,\uxi,\umu,\uSigma)-p(y;\ux,\uxi^{\prime},\umu,\uSigma)|+ \nonumber\\
& \text{N}(\ux;\umu^{\prime},\uSigma^{\prime})|p(y;\ux,\uxi^{\prime},\umu,\uSigma)-p(y;\ux,\uxi^{\prime},\umu^{\prime},\uSigma^{\prime})|. \label{lim3}
\end{align}
Due to the equicontinuity of the multivariate normal pdf \citep{wu2008, canale2017} the first and last terms in the right-hand side (\ref{lim3})
can be made arbitrarily small for all $\uz \in \mathcal Z$. Furthermore, the middle term can be made arbitrarily small because 
of the following expression for the difference of the probabilities
\begin{align*}
&|p(y;\ux,\uxi,\umu,\uSigma)-p(y;\ux,\uxi^{\prime},\umu,\uSigma)| = \\
&\left|
\int_{\Phi^{-1}(F\{y-1;\uxi\})}^{\Phi^{-1}(F\{y;\uxi\})} \text{N}(y^*;m^{*},v^{*})dy^* - 
\int_{\Phi^{-1}(F\{y-1;\uxi^{\prime}\})}^{\Phi^{-1}(F\{y;\uxi^{\prime}\})} \text{N}(y^*;m^{*},v^{*})dy^*\right|,
\end{align*}
and the equicontinuity of the cut-point function $\Phi^{-1}(F\{y;\uxi\})$, viewed as a function of $\uxi$.  

Hence, for any $\delta > 0$, there exist $\uz_1,\ldots,\uz_m$, such that for any $\uz \in \mathcal Z$
\begin{align}
\sup_{\utheta \in \uTheta}|k(\uz;\utheta) - k(\uz_i;\utheta)|, \label{n1} 
\end{align}
for some $\uz_i, i=1,\ldots,m$. 

Let
\begin{align*}
\mathcal{U} = \{P: |\int_{\uTheta} k(\uz_i;\utheta) dP_m(\utheta) - \int_{\uTheta} k(\uz_i;\utheta) dP(\utheta)|< c \delta, i=1,\ldots,m\}.
\end{align*}
It follows that $\mathcal{U}$ is a weak neighbourhood of $P_m$ with $\Pi(\mathcal{U})>0$.

Now, for some $P \in \mathcal{U}$ and any $\uz \in \mathcal Z$ we have that
\begin{align}
& \left| f_p(\uz) - f_{p_m}(\uz) \right| = 
\left| \int k(\uz;\utheta) dP(\utheta) - \int k(\uz;\utheta) dP_m(\utheta) \right| \nonumber\\
& = \Big| \int k(\uz;\utheta) dP(\utheta) \pm \int k(\uz_i;\utheta) dP(\utheta) \nonumber\\
&\pm 
\int k(\uz_i;\utheta) dP_m(\utheta) - \int k(\uz;\utheta) dP_m(\utheta) \Big|, \label{abc}
\end{align}
where $\uz_i$ is chosen from (\ref{n1}) and with $\pm a$ we mean add and subtract $a$.  

It follows that the expression (\ref{abc}) is $\leq 3 c \delta$. Further, recalling (\ref{elev}), from which follows that
$f_{p_m}(\uz) = \int k(\uz;\utheta) dP_m(\utheta) > c$, and dividing both sides of (\ref{abc}) 
by $f_{p_m}$, we obtain
\begin{align*}
\left| \frac{f_{p_m}(\utheta)}{f_{p}(\utheta)} - 1 \right| \leq \frac{3 \delta}{1-3\delta}.
\end{align*}

Hence, condition $A_2$ is satisfied for any $P \in \mathcal{U}$ as
\begin{align*}
\sum_y \int f_0(y,\ux) \log\frac{f_{P_{m}}(y,\ux)}{f_P(y,\ux)}d\ux < \frac{3 \delta}{1-3\delta}.
\end{align*}
This completes the proof. 

\newpage
\bibliographystyle{anzsj}
\bibliography{all}
\end{document}